\documentclass[fleqn,usenatbib,useAMS]{mnras}

\usepackage{graphicx}	% Including figure files
\usepackage{amsmath}	% Advanced maths commands
\usepackage{amssymb}	% Extra maths symbols
\usepackage{multicol}   % Multi-column entries in tables
\usepackage{bm}		    % Bold maths symbols, including upright Greek
\usepackage{tabularx}

\newcommand{\kms}{\,km\,s$^{-1}$} % kilometres per second
\def\alphaR{\gamma}
\def\alphaM{\alpha}
\def\Rp{R_{\rm p}}
\def\Ra{R_{\rm a}}

\def\rhoh{\rho_{\rm h}}
\def\rhoratio{\mathcal{R}}
\def\rhoration{\mathcal{R}_0}
\def\rhoratiocrit{\mathcal{R}_{\rm cr}}

\def\rhojeff{\rho_{\rm J,eff}}
\def\rhohn{\rho_{\rm h,0}}
\def\rhohf{\rho_{\rm h,f}}

\def\Mlo{M_{\rm lo}}
\def\Mup{M_{\rm up}}
\def\Rlo{R_{\rm lo}}
\def\Rup{R_{\rm up}}

\def\trh{\tau_{\rm rh}}
\def\trhn{\tau_{\rm rh,0}}
\def\rh{r_{\rm h}}
\def\rhn{r_{\rm h,0}}
\def\rj{r_{\rm J}}
\def\rjeff{r_{\rm J,eff}}

\def\omegatid{\Omega_{\rm tid}}

\def\dr{{\rm d}}

\def\kms{{\rm km/s}}

\def\myr{{\rm Myr}}
\def\gyr{{\rm Gyr}}
\def\pc{{\rm pc}}

\def\kpc{{\rm kpc}}

\def\msun{{\rm M}_\odot}
\def\msunmyr{{\rm M}_\odot\,{\rm Myr}^{-1}}
\def\msunpc{{\rm M}_\odot\,{\rm pc}^{-3}}

\def\Mbh{M_{\rm BH}}

\def\Mbhdot{\dot{M}_{\rm BH}}

\def\Ms{M_{\rm c}}

\def\fbh{f_{\rm BH}}
\def\fbhn{f_{\rm BH,0}}

\def\fsev{\mu_{\rm sev}}

\def\mref{M_{\rm ref}}
\def\mdotref{\dot{M}_{\rm ref}}

\def\mto{M_{\rm TO}}
\def\NGC{N_{\rm GC}}
\def\fsurv{f_{\rm surv}}
\def\Vc{V_{\rm c}}
\def\Vr{V_{\rm r}}
\def\sigr{\sigma_{\rm r}}
\def\sigt{\sigma_{\rm t}}

\def\Vt{V_{\rm t}}
\def\rg{R}
\def\rgvec{\boldsymbol{\rg}}
\def\vgvec{\boldsymbol{V}}
\def\rgeff{R_{\rm eff}}
\def\rga{R_{\rm ani}}
\def\feh{{\rm [Fe/H]}}
\def\NGC{N_{\rm GC}}

\def\Mps{M_{\rm i}}
\def\mmin{M_{\rm i, min}}

\def\Lgal{L_{\rm gal}}

\def\ecc{\epsilon}

\def\slogm{\sigma_{\log_{10}\!M}}
\def\taum{\tau_M}
\def\tdis{t_{\rm dis}}

\def\fmp{f_{\rm MP}}
\def\mulog{\mu_{\log_{10}\!M}}
\def\siglog{\sigma_{\log_{10}\!M }}

\usepackage[T1]{fontenc}
\usepackage{ae,aecompl}

\usepackage{txfonts}
\usepackage{soul}
\usepackage[normalem]{ulem}
\usepackage{soul}
\usepackage[dvipsnames]{xcolor}

\newif\ifnew
\newtrue
\def\bfnew#1{#1}

\def\mgnew#1{}

\ifnew
\def\bfnew#1{{\bfnew{#1}}}

\def\mgnew#1{{\it\textcolor{red}{\,MG:#1}}}
\fi
\definecolor{orange}{rgb}{1, 0.35, 0.}

\title[The mass-loss rates of star clusters with BHs]
{The mass-loss rates of star clusters with stellar-mass black holes: implications for the globular cluster mass function}
\author[]{Mark Gieles$^{1,2}$\thanks{Contact e-mail:\href{mailto:mgieles@icc.ub.edu}{mgieles@icc.ub.edu}}\href{https://orcid.org/0000-0002-9716-1868}{\includegraphics[scale=0.5]{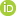}},
Oleg Y. Gnedin$^{3}$\href{https://orcid.org/0000-0001-9852-9954}{\includegraphics[scale=0.5]{orcid.png}}\\
$^1$ICREA, Pg. Llu\'{i}s Companys 23, E08010 Barcelona, Spain\\
$^2$Institut de Ci\`{e}ncies del Cosmos (ICCUB), Universitat de Barcelona (IEEC-UB), Mart\'{i} i Franqu\`{e}s 1, E08028 Barcelona, Spain\\
$^{3}$Department of Astronomy, University of Michigan, Ann Arbor, MI 48109, USA\\
}

\date{Accepted 2023 April 25. Received 2023 April 25; in original form 2023 March 7}

\pubyear{2023}

\begin{document}
\label{firstpage}
\pagerange{\pageref{firstpage}--\pageref{lastpage}}
\maketitle

\begin{abstract}  % MAX 250 words, currently 231 words
Stellar-mass black holes (BHs) can be retained in globular clusters (GCs) until the present. Simulations of GC evolution find that the relaxation driven mass-loss rate is elevated if BHs are present, especially near dissolution. We capture this behaviour in a parameterised mass-loss rate, benchmarked by results from $N$-body simulations, and use it to evolve an initial GC mass function (GCMF), similar to that of young massive clusters in the Local Universe, to an age of 12 Gyr. Low-metallicity GCs ($\feh\lesssim-1.5$) have the highest mass-loss rates, because of their relatively high BH masses, which combined with their more radial orbits and  stronger tidal field in the past explains the high turnover mass of the GCMF ($\sim10^5\,\msun$) at large Galactic radii ($\gtrsim 10\,\kpc$). The turnover mass at smaller Galactic radii is similar because of the upper mass truncation of the initial GCMF and the lower mass-loss rate due to the higher metallicities. The density profile in the Galaxy of mass lost from massive GCs ($\gtrsim10^{5}\,\msun$) resembles that of nitrogen-rich  stars in the halo, confirming that these stars originated from GCs. We conclude that two-body relaxation is the dominant effect in shaping the GCMF from a universal initial GCMF, because including the effect of BHs reduces the need for additional disruption mechanisms. 
\end{abstract}
\begin{keywords}
galaxies: star clusters: general -- globular clusters: general -- stars: black holes
\end{keywords}

%%%%%%%%%%%%%%%%%%%%%%%%%%%%%%%%%%%%%%%%%%%%
\section{Introduction}

Globular cluster (GC) systems in the Milky Way and external galaxies have peaked logarithmic mass and luminosity functions, with a typical luminosity $M_V\simeq -7.5$ \citep[for example,][]{2001stcl.conf..223H,2007ApJS..171..101J}, corresponding to a peak mass $\sim2\times10^5\,\msun$ and a dispersion $\slogm\simeq0.5$. 
This is markedly different from young star clusters, which form with a power-law mass function with a slope of about $-2$ \citep*[see][for reviews]{2010ARA&A..48..431P,2019ARA&A..57..227K}.
Old GCs may have formed with a similar mass function, because low-mass clusters had time to dissolve as the result of various disruptive effects, such as two-body relaxation, tidal shocks and interaction with dense molecular gas clouds. This disruption could turn over a power-law initial GC mass function (GCMF) and impose a typical mass scale of $\sim10^5\,\msun$ in the surviving GCs \citep[for example,][]{1995MNRAS.274...48O,2001ApJ...561..751F,2008ApJ...689..919P,2010ApJ...712L.184E,2015MNRAS.454.1658K}.
Adopting the hypothesis that the physics of cluster formation in giant molecular clouds is similar at all cosmic times \citep{harris_pudritz94,elmegreen_efremov97}, several recent studies have confirmed that massive star clusters formed in high-redshift galaxies would evolve into old clusters matching the age-metallicity distribution and the spatial and kinematic distributions of observed GC systems \citep{choksi_etal18,pfeffer_etal18,kruijssen_etal19a, 2023MNRAS.521..124R}. However, the resulting GCMF in these models tends to be  skewed towards lower masses compared to the observed GCMF. Reproducing the shape of the GCMF is, therefore, one of the last remaining hurdles to confirm that star cluster formation is a universal mechanism in all epochs and environments.

Some studies have suggested that  GCs had a typical mass scale imprinted at formation \citep[for example,][]{1968ApJ...154..891P,1985ApJ...298...18F,2002ApJ...566L...1B,2016ApJ...823...52K}, and indeed an initially peaked mass distribution would preserve its shape as clusters lose mass \citep{2000MNRAS.318..841V,2001ApJ...561..751F}. However, such scenarios rely on physical conditions in the galactic interstellar medium that are unlikely to produce giant molecular clouds massive and dense enough to host proto-GCs \citep[e.g.,][]{forbes_etal18}.
We therefore prefer the hypothesis that the initial GCMF is universal across cosmic time and that disruption is responsible for the current shape of the GCMF.

The relative contribution of various disruption mechanisms is still debated. Because of the high rate of close stellar encounters in GCs, it is natural first to explore the effect of two-body relaxation in the large-scale galactic tidal field (hereafter, `evaporation'). 
The mass-loss rate due to evaporation depends on the strength of the tidal field \citep[for example,][]{1987ApJ...322..123L,1990ApJ...351..121C,2003MNRAS.340..227B}, and therefore it predicts an anti-correlation between the turnover mass ($\mto$) and galactocentric radius ($\rg$) for clusters in a static galactic potential and with constant velocity anisotropy. 
In contrast, the observed $\mto$ varies only mildly with $\rg$ in the Milky Way \citep[see, for example, figure 8 in][]{2008ApJ...679.1272M} and is also remarkably constant across galactic environments \citep{2007ApJS..171..101J}. We refer to this tension between model predictions and the observed near-universality of the GCMF as `the GCMF problem'.

Various studies have attempted to resolve the GCMF problem by either assuming a strong radially-biased velocity anisotropy  at large Galactic radii \citep[for example,][]{2001ApJ...561..751F}; invoking additional universal disruptive effects, such as stellar evolution \citep{2003ApJ...587L..97V} and gas expulsion \citep*{2008MNRAS.384.1231B}; or assuming that the mass-loss rate depends mainly on present-day density \citep{2008ApJ...679.1272M}. However, these assumptions are all in tension with results from observations \citep{2003ApJ...593..760V, 2019MNRAS.484.2832V}, theory \citep*{1961AnAp...24..369H,2011MNRAS.413.2509G}, and numerical simulations of cluster evolution in tidal fields \citep{1987ApJ...322..123L,2003MNRAS.340..227B,2008MNRAS.389L..28G}.

In recent years, much attention has been given to the disruption by tidal interactions with giant molecular clouds in the first $\sim1\,\gyr$ \citep{2010ApJ...712L.184E,2015MNRAS.454.1658K,pfeffer_etal18}. Tidal shocks preferentially destroy low-density clusters \citep*{1958ApJ...127...17S,1972ApJ...176L..51O} and therefore not necessarily low-mass clusters, but relaxation leads to an expansion of low-mass clusters, reducing their densities, such that the combined effect of relaxation and tidal perturbations leads to a mass-dependence of the disruption timescale that is similar to that of evaporation \citep{2016MNRAS.463L.103G}.

Most cluster population studies mentioned above rely on prescriptions for evaporation based on theory of equal-mass clusters by \citet{1961AnAp...24..369H}, or results of numerical $N$-body simulations of clusters with a stellar mass function and stellar evolution \citep[for example,][]{2003MNRAS.340..227B}, but without stellar-mass black holes (BHs). 
However, BH candidates have been reported in several Milky Way GCs \citep{2012Natur.490...71S, 2013ApJ...777...69C,2015MNRAS.453.3918M,2018MNRAS.475L..15G, 2020A&A...635A..65K} and in extra-galactic clusters \citep{2007Natur.445..183M, 2011MNRAS.410.1655M,2012ApJ...757...40B,2022MNRAS.511.2914S}. 

These discoveries of BHs in clusters led to various modelling efforts of GCs with BHs, that showed that clusters in a tidal field dissolve faster if they retain a significant fraction of their BHs after natal kicks  \citep{2017ApJ...834...68C,2017MNRAS.470.2736P,2019MNRAS.487.2412G,2020MNRAS.491.2413W, 2020ApJS..247...48K,2021NatAs...5..957G}. Apart from shortening the total lifetime, the mass evolution over time is also different, in the sense that the (absolute) mass-loss rate increases towards dissolution. This is because tidally limited clusters with a BH mass fraction at a critical value of a few percent will lose BH mass at the same rate as stellar mass and therefore maintain that constant BH mass fraction \citep{2013MNRAS.432.2779B}. If the mass fraction in BHs is higher(lower), the BH fraction continues to increase(decrease) \citep{2011ApJ...741L..12B, 2021NatAs...5..957G}. An increasing BH fraction with time leads to an increasing (absolute) mass-loss rate and such an abrupt dissolution leads to a concave shape of the mass evolution with time $M(t)$ \citep[][]{2019MNRAS.487.2412G}. In contrast, for clusters without BHs the shape of $M(t)$ is convex \citep{2001MNRAS.325.1323B,2008MNRAS.389L..28G}. We will refer to the  concave and convex shapes of $M(t)$  as `jumping' and `skiing', respectively, following terminology from \citet*{2015MNRAS.449L.100C}. 
In this work we propose an analytical prescription for the mass-loss rate that allows for different shapes of $M(t)$, with a flexible dependence of the total lifetime on the initial mass, informed by results of a grid of direct $N$-body simulations. We then use it to model the evolution of the GCMF in a Milky Way-like galaxy. 

This paper is organised as follows. In Section~\ref{sec:nbody} we analyse the mass-loss rate  in $N$-body simulations of star clusters with BHs. 
In Section~\ref{sec:mdot} we parameterise the mass-loss rate guided by the $N$-body simulations. We present a model for the evolution of the GCMF in Section~\ref{sec:model}, present the results in Section~\ref{sec:results} and then discuss broader implications of our results in Section~\ref{sec:discussion}. Our conclusions are summarised in Section~\ref{sec:conclusions}.

%%%%%%%%%%%%%%%%%%
\section{Insight from \textit{N}-body models}
\label{sec:nbody}

%____________________________________
\subsection{Description of the models}
\label{ssec:models}

To quantify the effect of BHs on the mass-loss rate ($\dot{M}$) of star clusters, we use the $N$-body models presented in \citet{2021NatAs...5..957G} that were performed with {\sc nbody6++gpu} \citep{2003gnbs.book.....A,2015MNRAS.450.4070W}. The grid of models in that work was intended to find the initial conditions of the Milky Way GC Palomar 5 (Pal 5, hereafter), hence all clusters are on the same orbit in a three-component Milky Way potential, with an apocentre distance $\approx 15.5\,\kpc$ and a pericentre distance $\approx 6.5\,\kpc$ (implying an orbital eccentricity $\epsilon\simeq 0.41$). We use the first 11 models from their table~1, which is a grid of models with different initial density within the half-mass radius ($\rhn$) of $\rhohn\equiv 3M_0/(8\pi\rhn^3)=\{30,100,300,1000\}\,\msunpc$ and number of stars $N=\{0.5,1,2\}\times10^5$, which for the adopted \citet{2001MNRAS.322..231K} stellar initial mass function (IMF) in the range $0.1-100\,\msun$ corresponds to initial cluster masses $M_0=\{0.32, 0.64,1.28\}\times10^5\,\msun$.  
The models adopt the rapid supernova mechanism  \citep{2012ApJ...749...91F} with the natal kicks lowered by the amount of fallback such that momentum is conserved. As a result, 63\% (73\%) of the number (mass) of BHs do not receive a natal kick for the adopted IMF and the metallicity of $Z=0.0006$ ($\feh\simeq -1.4$, using $Z_\odot=0.014$ for the solar metallicity, \citealt{2021A&A...653A.141A}).
The model with $\rhohn=10^3\,\msunpc$ and $N=2\times10^5$ was not run for that study, and we run it here with the same settings as the other models. We also run here two additional models with higher metallicity: $Z = 0.006~(\feh\simeq-0.4$) and $Z = 0.017~(\feh\simeq0)$, both with $N=10^5$ and $\rhohn=300\,\msunpc$. 

We compare the mass-loss rates of these models to the frequently-cited $N$-body models of \citet{2003MNRAS.340..227B}. These models consider a galactic tidal field due to a singular isothermal sphere (SIS) with circular velocity $\Vc=220\,\kms$, the effects of stellar evolution, and a stellar IMF truncated at $15 \,\msun$ such that no BHs form. Their mass-loss rates can be approximated as \citep*{2013MNRAS.433.1378L,choksi_etal18}
\begin{equation}
\dot{M} \simeq -30\,\msunmyr\,\left(\frac{M}{2\times10^5\,\msun}\right)^{1/3} \frac{\omegatid}{{0.32}\,\myr^{-1}}.
\label{eq:bm03}
\end{equation}
The influence of the tidal field is captured by $\omegatid$, 
which depends on the tidal and centrifugal  forces. For circular orbits it can be expressed through the first and third eigenvalues of the tidal tensor as $\omegatid = \sqrt{\lambda_1-\lambda_3}$ \citep*{2011MNRAS.418..759R,chen_gnedin23}. For the SIS $\lambda_1=-\lambda_3 = \Vc^2/\rg^2$ such that $\omegatid = \sqrt{2}\Vc/\rg$.
For eccentric orbits we use $\omegatid=\sqrt{2}\Vc/\rgeff$, where $\rgeff$ is the radius of the circular orbit with the same life time.
\citet{2003MNRAS.340..227B} show that for relaxation driven mass-loss of clusters in a SIS this effective radius is $\rgeff\equiv\Rp(1+\epsilon)=\Ra(1-\epsilon)$, where $\Rp$ and $\Ra$ are the pericentre and apocentre distance of the orbit, respectively \citep[see also][]{2016MNRAS.455..596C}. We normalise $\omegatid$ to a value corresponding to $\Vc=220\,\kms$ and $\rgeff=1\,\kpc$. 

For clarity, in this work we also adopt the SIS to approximate the Galaxy potential. The Pal 5 models did not evolve in a SIS, but in a more realistic three-component Milky Way, so we can use those results as an anchor point from which we extrapolate to larger and smaller $\rgeff$ by using the simple analytic properties of the SIS.
For the Pal 5 models $\rgeff \simeq 9.15\,\kpc$ and therefore {$\omegatid\simeq0.035\,\myr^{-1}$}. The density within the effective Jacobi radius ($\rjeff$) for a SIS is given by $\rhojeff = 3/(2\pi G)\omegatid^2$, where $G$ is the gravitational constant. For the models on the orbit of Pal 5 it is $\rhojeff\simeq 0.064\,\msunpc$. 

To describe the dependence of our results on the initial cluster density, we introduce a dimensionless parameter $\rhoratio\equiv\rhoh/\rhohf$, where $\rhohf$ is the half-mass density of a Roche-filling cluster. Roche filling in the context of clusters is ill-defined, but here we take it as a cluster that has $\rh/\rj=0.145$, which is the filling factor in the tidally limited cluster on a circular orbit of \citet{1961AnAp...24..369H}. The Pal 5 models have $\rhohf = 0.5\times0.145^{-3}\,\rhojeff\simeq10.5\,\msun\pc^{-3}$.
The relation between density and filling factor in terms of radius is $\rh/\rjeff = 0.145\rhoratio^{-1/3}$. Table~\ref{tab:rho} relates $\rhohn$ of the $N$-body models to these more physically-relevant quantities.

\begin{figure}
\includegraphics[width=\columnwidth]{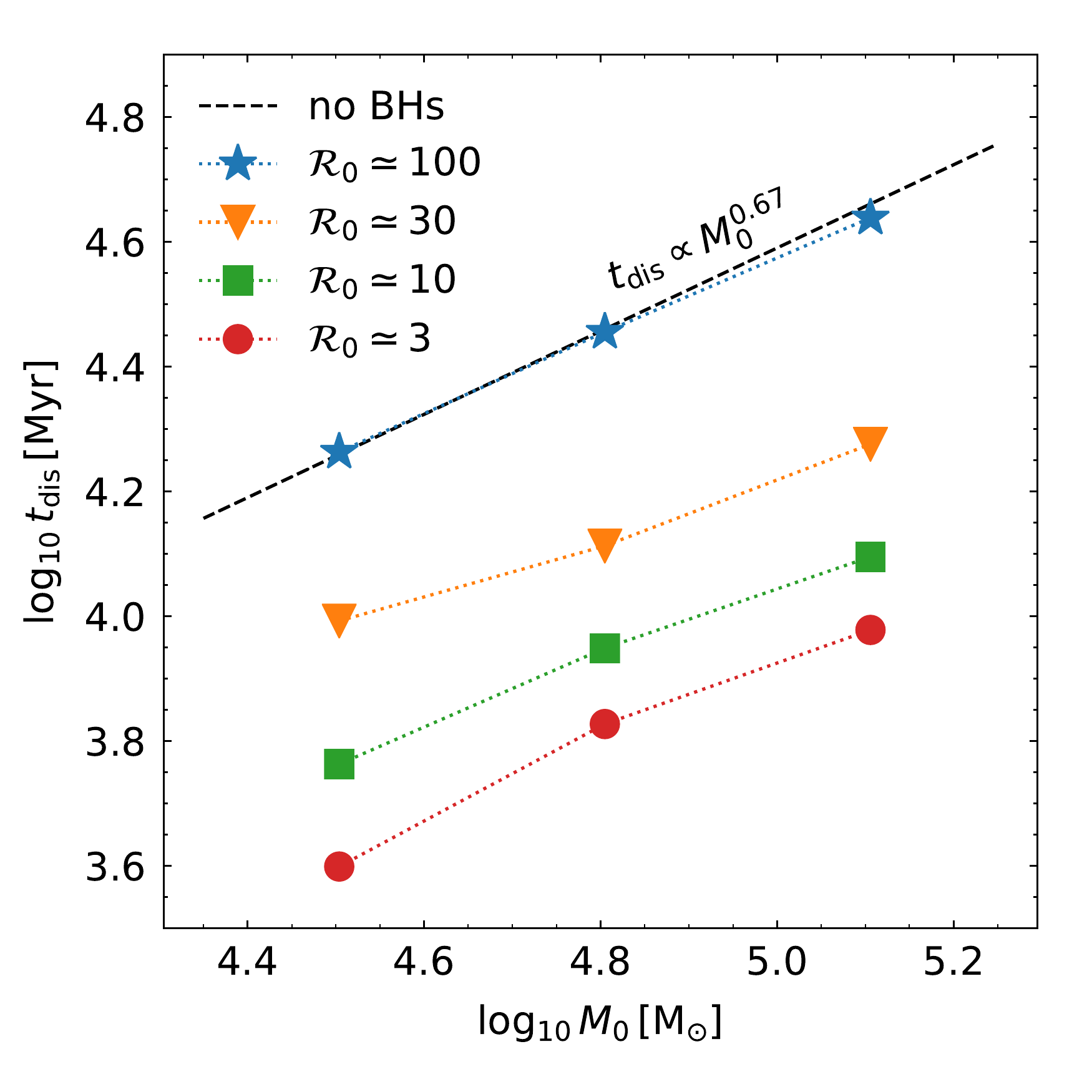}
\vspace{-5mm}
\caption{Disruption times ($\tdis$) for different $M_0$ and $\rhoration$ (see Table~\ref{tab:rho}) of all 12 $N$-body models on the orbit of Pal 5 from \citet{2021NatAs...5..957G}. Dashed line shows the expected $\tdis$ for models without BHs, as derived from equation~(\ref{eq:bm03}), see text for detail. The densest $N$-body models ($\rhoration=100$) have similar $\tdis$ because they eject all their BHs dynamically (Section~\ref{ssec:mdot}). Lower density clusters dynamically retain a BH population until dissolution and disrupt up to 4.5 times faster.}
\label{fig:tdis}
\end{figure}

In Fig.~\ref{fig:tdis} we show the disruption times ($\tdis$) of all 12 $N$-body models with low metallicity. We define $\tdis$ as the time needed for the mass to reach zero, although in the $N$-body models we only determine the time when the cluster reaches a low enough mass of $300\,\msun$. The dashed line shows $\tdis$ following from $\dot{M}$ for models without BHs (equation~\ref{eq:bm03}), which for this mass dependence of $\dot{M}$ is $\tdis(M_0) = 1.5\Mps/|\dot{M}(\Mps)|$ \citep{2005A&A...441..117L}, where 
\begin{equation}
\Mps \equiv \fsev M_0
\label{eq:Mps}
\end{equation}
is the initial mass after (most) stellar evolution related mass loss has occurred. From hereon we make the simplifying assumption that stellar evolution happens independently from evaporation, which is justified by the different timescales on which they operate (several $10\,\myr$ for most stellar evolution mass loss to occur vs. several Gyr for evaporation).
Here $\fsev\simeq0.55$ is the remaining mass fraction after stellar evolution for for a metallicity of $Z=0.0006$ $(\feh\simeq-1.4)$. Expressing $\tdis$ in terms of the initial mass after stellar evolution is needed because equation~(\ref{eq:bm03}) only describes the mass loss due to evaporation. This prediction for $\tdis$ agrees well with the results from the densest clusters, while the clusters with $\rhoration\simeq\{30,10,3\}$ dissolve approximately a factor of $\{2,3,4.5\}$ faster. In the next section we discuss in more detail the role of BHs in this trend of $\tdis(\rhoration)$.

\begin{table}
    \centering
    \begin{tabular}{ccccc}
        \hline\\
        $\rhohn$ & $\displaystyle\rhoration=\frac{\rhohn}{\rhohf}$ & $\displaystyle\frac{\rhn}{\rjeff}$ & $\mdotref$ & $y$\\
        $[\msun\,\pc^{-3}]$ &  &  & [$\msun\,\myr^{-1}$]\\
        \hline
        30 & 2.9 & 0.10 & $-95$ & 2\\
        100 & 9.5 & 0.068 & $-60$ & 1.75\\
        300 & 29 & 0.047 & $-45$ & 1.33\\
        1000 & 95 & 0.032  & $-30$ & 0.67\\
        \hline
    \end{tabular}
    \caption{Parameters of the $N$-body models shown in Fig.~\ref{fig:tdis}. For each initial density, models with $N=(0.5,1,2)\times10^5$ stars were run, corresponding to initial masses of $M_0\simeq(0.32,0.64,1.28)\times10^5\,\msun$. All models adopt a metallicity of $Z=0.0006~(\feh\simeq-1.4)$ and two additional models with higher metallicity were run for $\rhoration= 29$. The parameters $\mdotref$ and $y$ in the last two columns are used to approximate $\dot{M}(M,M_0)$ (equation~\ref{eq:mdot}) in Fig.~\ref{fig:mdot_nbody} (dashed line, top row).}
    \label{tab:rho}
\end{table}

%____________________________________
\subsection{Mass-loss rates}
\label{ssec:mdot}

\begin{figure*}
\includegraphics[width=\textwidth]{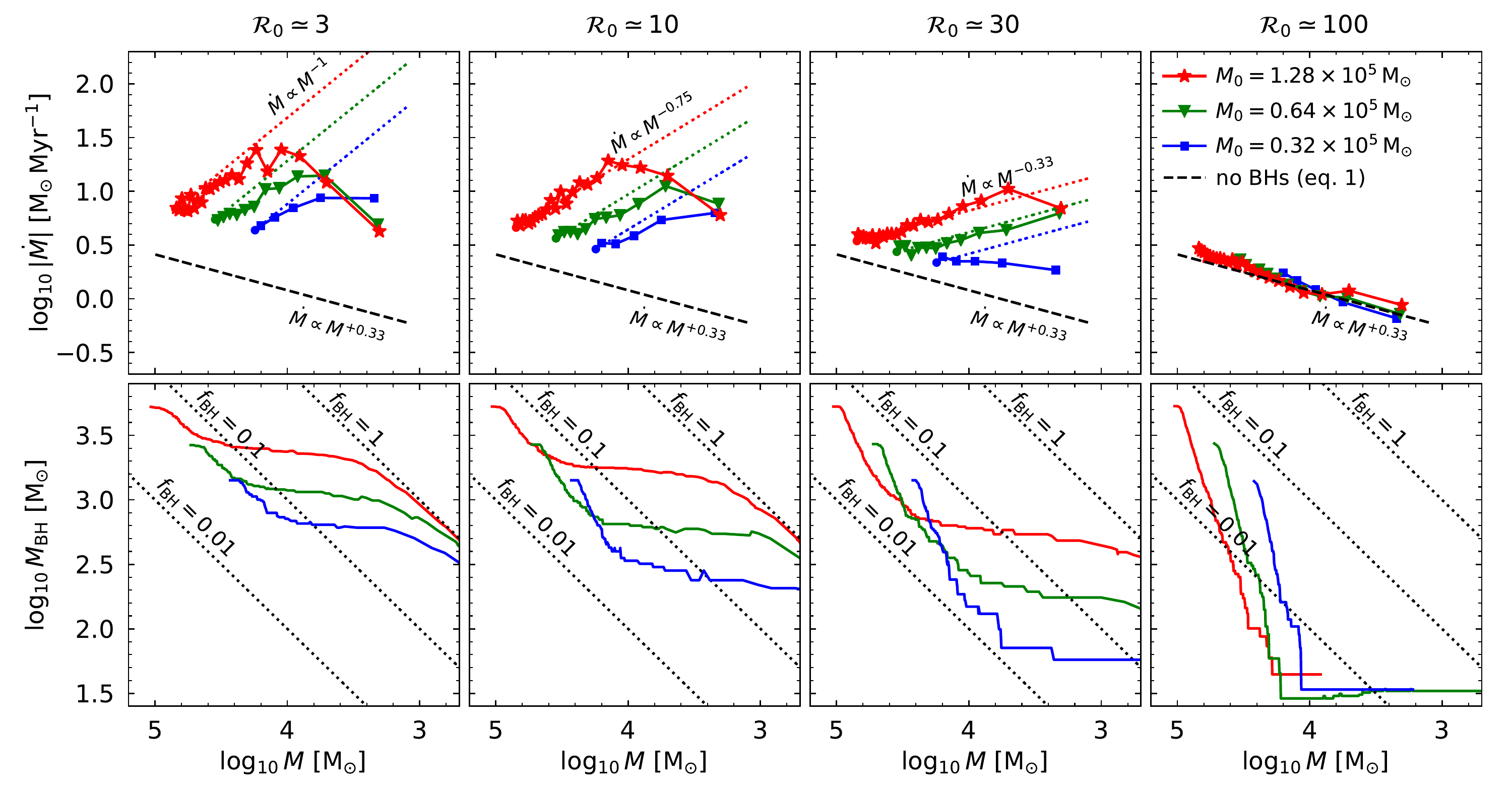}
\vspace{-5mm}
\caption{\textit{Top:} Mass-loss rates of $N$-body models on the orbit of Pal 5 from \citet{2021NatAs...5..957G}. Current cluster mass $M(t)$ acts as a proxy for time, flowing from left to right. Dotted lines show $\dot{M}$ curves from equation~(\ref{eq:mdot}) that match the $N$-body points. The dashed line is the same in every top panel and shows $\dot{M}$ without the effect of BHs. \textit{Bottom:} Mass in BHs in the same $N$-body models. Diagonal dotted lines mark constant BH fractions of 1\%, 10\%, and 100\%.
}
\label{fig:mdot_nbody}
\end{figure*}

In this section we describe how $\dot{M}$ depends on the initial conditions of the clusters. Because all clusters lose about 45\% of their initial mass by stellar evolution, mostly in the first Gyr, and we are here interested in evaporation, we determine $\dot{M}$ in the range $500\,\msun\le M <\Mps$ in mass bins with widths of $3\times10^3\,\msun$. In Fig.~\ref{fig:mdot_nbody} we show $\dot{M}$ from the $N$-body models in the top row, with different $M_0$ (different colours and symbols) and different  initial densities (different columns). The bottom row shows the remaining mass in BHs ($\Mbh$). The clusters with relatively low densities (left two columns) keep a significant fraction of their BHs and the mass-loss rate of these models increases towards dissolution.

There is a clear trend for higher density clusters to lose more of their BHs, which is the result of their shorter relaxation time \citep{2013MNRAS.432.2779B}.
The densest clusters ($\rhoration\simeq100$, right column) eject almost all BHs early and evolve along similar tracks as models without BHs (equation~\ref{eq:bm03}), shown as black dashed lines in the top row. This is why their $\tdis$ is similar to those of clusters without BHs (Fig.~\ref{fig:tdis}). \citet{2013MNRAS.432.2779B} explain that for tidally limited clusters there exists a critical $\fbh\simeq0.1$ at which the fraction of the total mass that is lost is in the form of BHs equals $0.1$, such that $\fbh$ remains constant. If $\fbh\lesssim0.1$ then all BHs are ejected, while if $\fbh\gtrsim0.1$ the cluster evolves towards a 100\% BH cluster \citep[see also][]{2011ApJ...741L..12B}. This was derived for idealised two-component models. In our models we find that this critical fraction is lower: $\fbh\simeq0.025$. This has consequences for clusters with higher metallicity, because they form with a lower $\fbh$ than metal-poor clusters and therefore drop more easily below the critical $\fbh\simeq0.025$ (Section~\ref{ssec:metallicity}).

The fact that the different coloured $\dot{M}$ points in each panel do not overlap shows that at the same remaining mass $M$, models with different $M_0$ have different $\dot{M}$, because their $\fbh$ are different. 
This behaviour is reproduced {for most parts of the evolution} by the dotted lines, which are a simple parameterisation of $\dot{M}$, described in more detail in Section~\ref{sec:mdot} (equation~\ref{eq:mdot}). They are power-law relations for $\dot{M}(M,M_0)$ of the form $\dot{M}\propto (M/\Mps)^a\Mps^{1/3}$, where the value of $a$ required to described (most of) the data ranges from $a=-1$ to $a=1/3$. 
This relation results in a dependence $\tdis\propto M_0^{2/3}$ (see Fig.~\ref{fig:tdis}) independent of the value of $a$, that is, the same $M_0$ dependence as was found for models without BHs (equation~\ref{eq:bm03}). 

For $\fbh\gtrsim0.3$, $|\dot{M}|$ decreases again, because then the stars become less important and the cluster evolves from a two-component model (stars and BHs) to a single-component model (only BHs). This leads to a narrower mass spectrum and a slower evolution, but still $\sim10$ times faster than for a cluster with only stars. In fact, $\dot{M}(M)$ is then evolving parallel to the dashed line shown in the top row of Fig.~\ref{fig:mdot_nbody}, but above it because of the higher mean  mass, which reduces $\trh$.  

\begin{figure}
\includegraphics[width=\columnwidth]{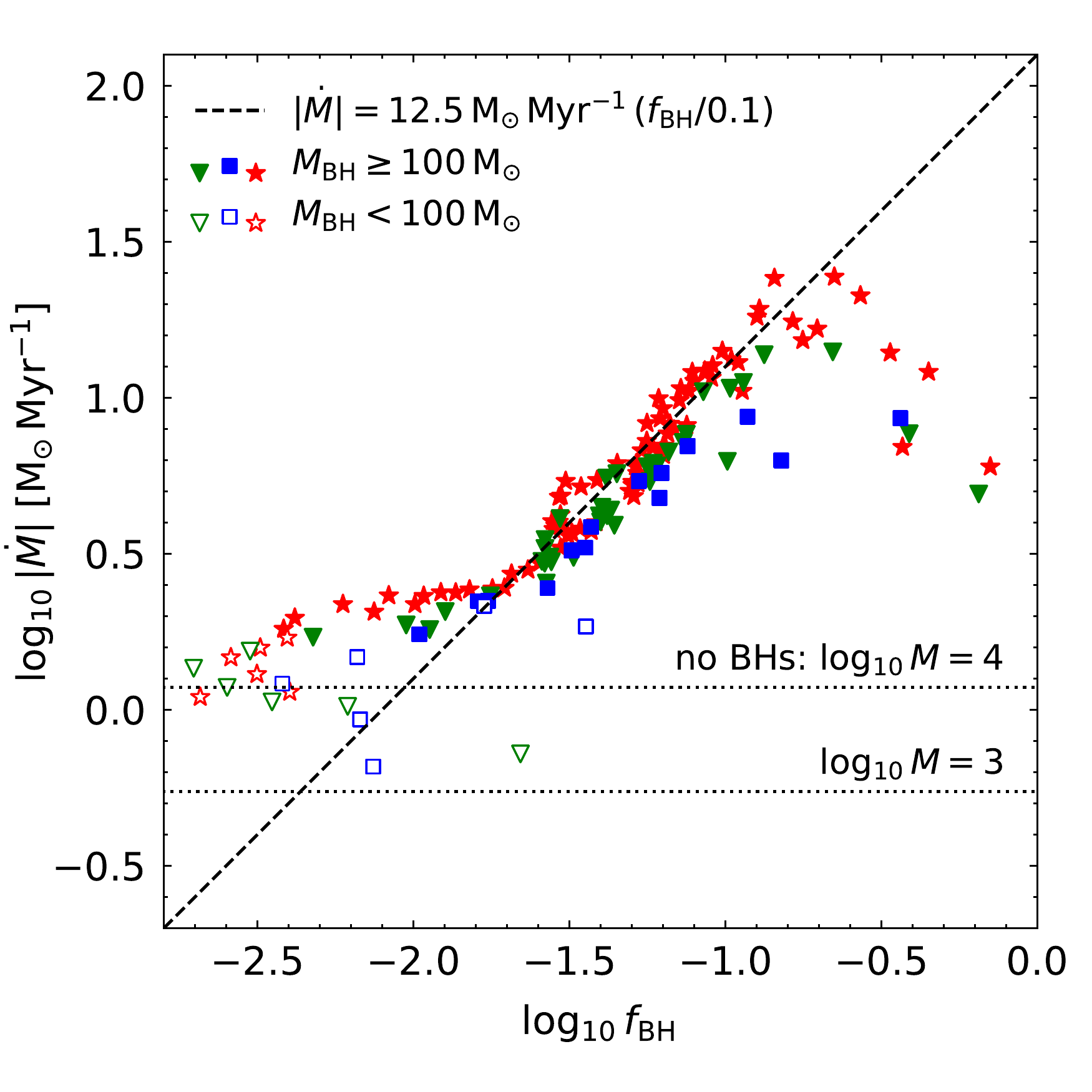}
\vspace{-8mm}
\caption{Mass loss rate at different $\fbh$ for all models with $\feh=-1.4$. We plot the models with $\Mbh<100\,\msun$ as open symbols, because these only have a few BHs, and these clusters are not expected to behave in the same way as the cluster with a population of BHs. Dashed line is a linear approximation for the relation $\dot{M}(\fbh)$.}
\label{fig:mdot_fbh}
\end{figure}

Fig.~\ref{fig:mdot_fbh} shows $\dot{M}$ as a function of the remaining BH fraction $\fbh$ for all low-metallicity models. The increase of $|\dot{M}|$ with $\fbh$ can be approximated by a linear relation $\dot{M} \simeq -12.5\,\msun\,\myr^{-1} \left(\fbh/0.1\right)$. As shown earlier, clusters evolve at roughly constant $\Mbh$ at late stages, such that a relation $\dot{M} \propto \fbh$ implies $\dot{M} \propto M^{-1}$. This leads to a strongly `jumping' $M(t)$, very different from the result of models without BHs ($\dot{M} \propto M^{1/3}$).

Clusters lose most of their BHs in the early expansion phase. Once the cluster density has become comparable to (some multiple of) the tidal density, the cluster evolves at approximately constant $\Mbh$ (see bottom row of Fig.~\ref{fig:mdot_nbody}). 
We can estimate how the BH loss in the expansion phase depends on the initial conditions. \citet{2013MNRAS.432.2779B} showed that the mass-loss rate of the BH population depends on the cluster properties as $\Mbhdot \propto M/\trh$, where $\trh$ is the half-mass relaxation timescale.  
We approximate the total BH mass lost as $\Delta \Mbh = \Mbhdot \Delta t$, where $\Delta t$ is the time the cluster needs to fill the Roche volume. In the expansion phase the density reduces in time as $\rho(t) \simeq \rhohn (t/\trhn)^{-2}$, where $\trhn$ is the initial $\trh$ \citep{1965AnAp...28...62H,2011MNRAS.413.2509G}.
So the time the cluster needs to expand to the tidal boundary is $\Delta t \simeq \trhn \rhoration^{1/2}$. 
Using also the initial values in the expression for $\Mbhdot$, we thus find $\Delta \Mbh \propto M_0\rhoration^{1/2}$ or $\Delta \fbhn \propto\rhoration^{1/2}$. So the reduction of $\fbh$ due to dynamical ejections depends only on the initial density, relative to the tidal density. 
The data in the bottom row of Fig.~\ref{fig:mdot_nbody} show that indeed that the drop in $\fbh$ is larger for the higher initial densities.
We also note that there is a small dependence on $M_0$ for $\rhoration=10-30$, with the drop in $\Mbh$ being (relatively) large for the low-mass clusters. This is in the regime where only a handful of BHs are left and therefore the theory of Breen \& Heggie no longer holds and we will not attempt to capture this. 

There exists a critical density $\rhoratiocrit$ between the models with $\rhoration=30$ and $\rhoration=100$ above which all BHs are ejected. We  propose a simple relation for the drop in $\fbh$ because of the dynamical ejection of the form
\begin{equation}
\Delta\fbh\propto\left(\frac{\rhoration}{\rhoratiocrit}\right)^{1/2},
\label{eq:dfbh}
\end{equation}
for $\rhoration<\rhoratiocrit\simeq50$. Metal-rich clusters have lower $\fbhn$, and because the constant of proportionality in equation~(\ref{eq:dfbh}) does not depend on $\fbhn$, the critical density is found from equating $\Delta \fbh=\fbhn$ such that  $\rhoratiocrit\propto\fbhn^2$.  
We discuss the consequences for metallicity next.

%_________________________
\subsection{Metallicity}
\label{ssec:metallicity}
\begin{figure}
\includegraphics[width=0.48\textwidth]{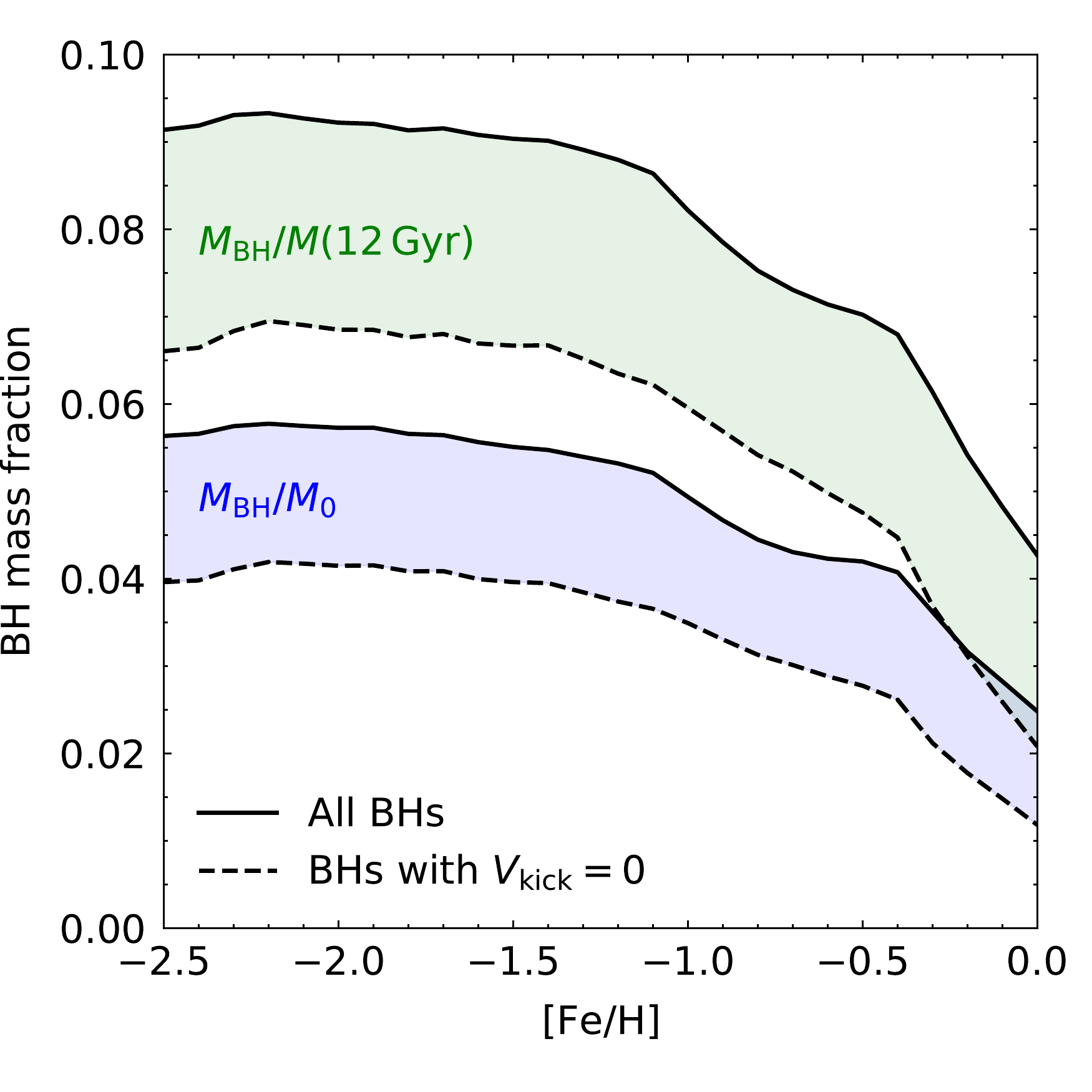}
\vspace{-6mm}
\caption{Mass fraction in stellar-mass BHs as a function of metallicity, from SSE \citep{2000MNRAS.315..543H,2020A&A...639A..41B}. BHs that receive no kick are shown by dashed lines, the sum of BHs with and without kick is shown by solid lines.}
\label{fig:fbh_feh}
\end{figure}

Stars of higher metallicity have stronger winds, which results in lower remnant masses. To quantify this effect on the masses of BHs, we adopt a \citet{2001MNRAS.321..699K} IMF in the range $0.1-100\,\msun$ and evolve the stars to an age of 12 Gyr with the single stellar evolution model SSE by \citet*[][]{2000MNRAS.315..543H}, with the recent update for massive star winds from \citet{2020A&A...639A..41B}. We then compute the mass fraction in BHs and show it in Fig.~\ref{fig:fbh_feh}, both in terms of the initial mass $(\fbhn=\Mbh/M_0$, blue shaded region) and in terms of the total mass at 12 Gyr $(\fbh=\Mbh/M$, green shaded region), which is roughly a factor $1/\fsev\simeq1.8$ higher. Since some of the BHs receive natal kicks {(see Section~\ref{ssec:models})}, we show the two extreme cases where all BHs that are kicked are either lost (dashed lines) or retained (full lines). As can be seen, $\fbh$ is roughly constant below $\feh\simeq-1.5$ (typical metal-poor GC) and decreases approximately by a factor of two going to $\feh\simeq-0.5$ (typical metal-rich GC).

\begin{figure}
\includegraphics[width=0.47\textwidth]{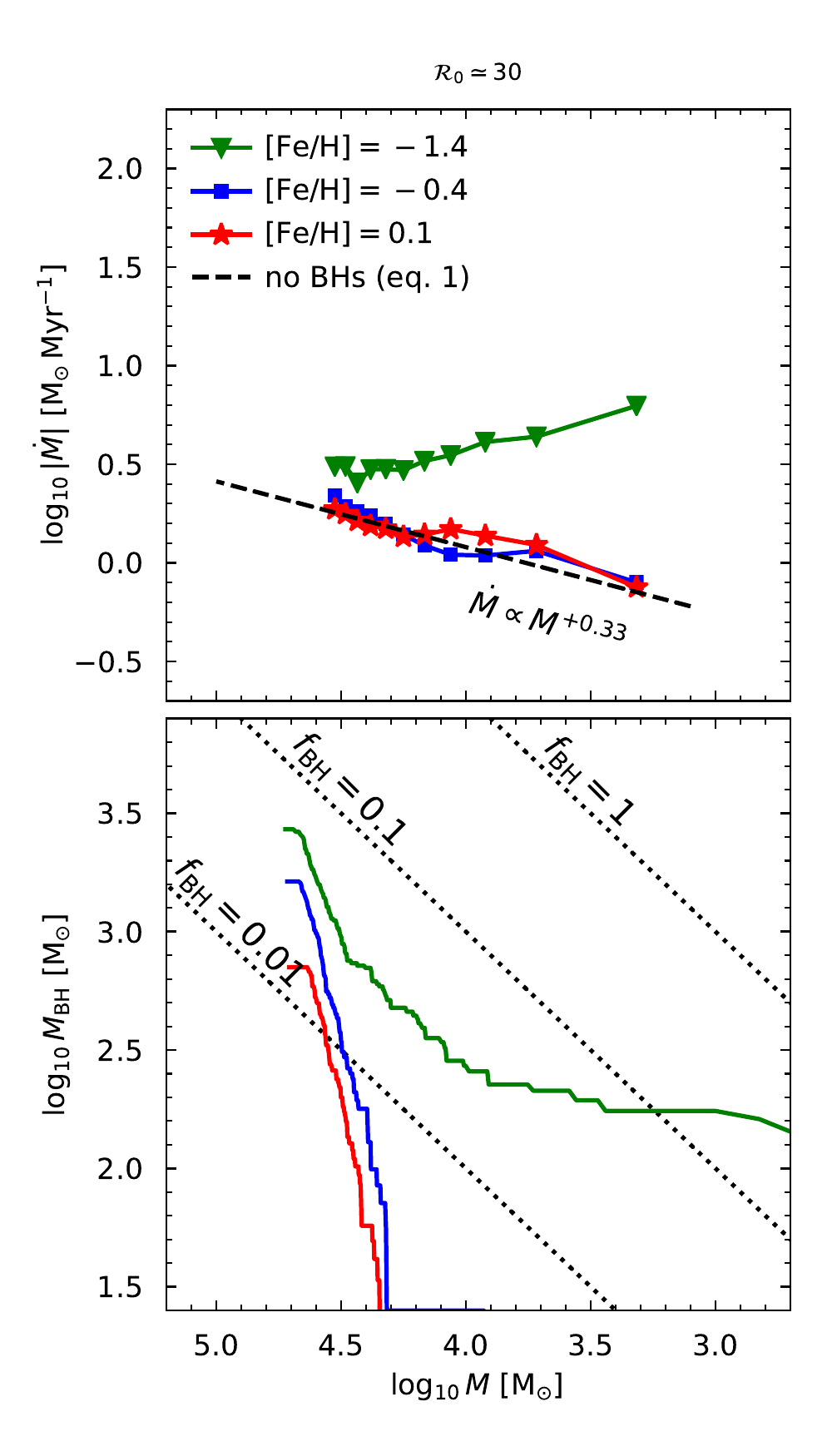}
\vspace{-5mm}
\caption{Results of $N$-body models with different initial metallicities. The model with $\feh=-1.4$ is the same as shown in Fig.~\ref{fig:mdot_nbody}. The models with higher $\feh$ have lower initial $\fbh$, leading to complete ejection of all BHs and mass-loss rates comparable to models without BHs (dashed line).}
\label{fig:z}
\end{figure}

To illustrate the effect of $\feh$ and the resulting $\fbh$ on the evolution of the cluster, we show here the results of two  $N$-body models with higher metallicity $Z = 0.006~(\feh\simeq-0.4)$ and $Z = 0.017~(\feh\simeq0.1)$ on the same orbit as the other models, and with $N=10^5$ and $\rhohn=300\,\msun\,\pc^{-3}\,(\rhoration\simeq30)$. For the metal-poor models shown in Fig.~\ref{fig:mdot_nbody}, sufficient number of BHs were retained for them to have a noticeable effect on $\dot{M}$ at this density. 
The effect of higher metallicity on $\fbh$ and $\dot{M}$ in $N$-body models is shown in Fig.~\ref{fig:z}. From this plot we see that the lower initial value of $\fbh$ results in all BHs being dynamically ejected and their $\dot{M}$ following the results of models without BHs (equation~\ref{eq:bm03}). Because $\fbhn$ is a factor of $\sim2$ lower at these higher metallicities, the critical initial density for ejection of all BHs is a factor of $\sim4$ lower (see text below equation~\ref{eq:dfbh}), so $\rhoratiocrit\simeq13$ instead of 50. This $\rhoratiocrit$ is now lower than the density of these models ($\rhoration\simeq30$) and explains why all BHs are ejected. We will use this metallicity dependence of $\dot{M}$ in the population model (Section~\ref{sec:model}).

\begin{figure*}
\includegraphics[width=0.95\textwidth]{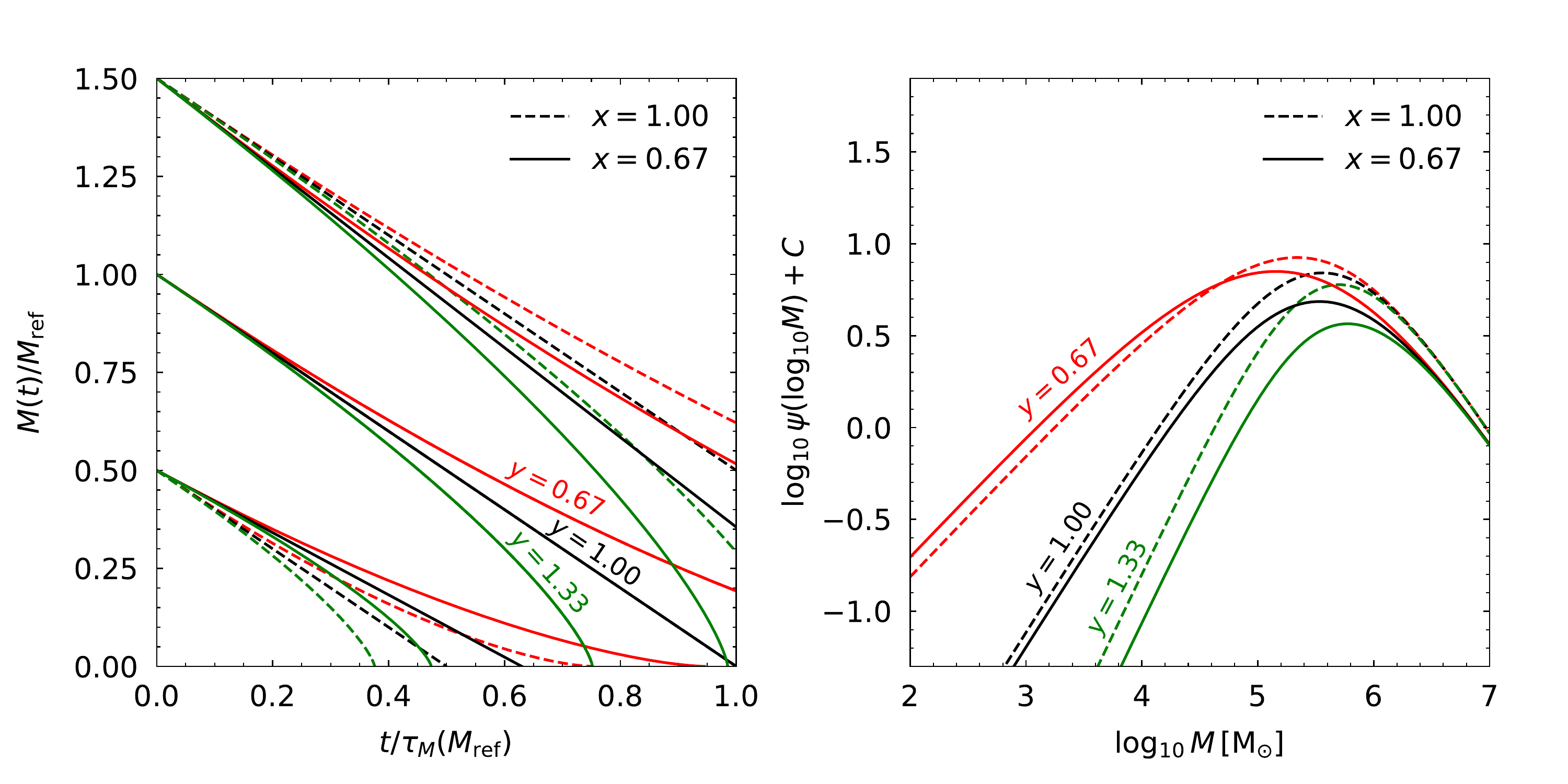}
\vspace{-2mm}
\caption{{\it Left}: Evolution of cluster mass for different $M_0$, $x$ and $y$. Here $\mref$ is a constant reference mass. {\it Right}: Evolved cluster mass function for the same set of $x$ and $y$ as in the left panel and for $\mdotref=-30\,\msun\,\myr^{-1}$ and $\omegatid=0.32\,\myr^{-1}$.}
\label{fig:gcmf0}
\end{figure*}

%%%%%%%%%%%%%%%%%%%%%%%%%%%%%%%%%%%%%%%%%
\section{A novel parameterisation of the cluster mass-loss rate}
\label{sec:mdot}

In this section we consider a simple analytical model for the evolution of the GCMF that results from adding the effects of BHs.

%_____________________________________
\subsection{Mass-loss rate and mass evolution}
If we assume clusters to be spherically symmetric and moving on circular orbits in a static galactic potential, the evaporation mass-loss rate would be independent of both cluster mass and time \citep[for example,][]{1961AnAp...24..369H, 2001ApJ...561..751F}. 
In this simple picture, the mass-loss timescale is always proportional to the current mass $M(t)$: $\taum \equiv -M/\dot{M}\propto M$. The total disruption time for a cluster with initial mass  $\Mps$ (that is, the mass after stellar evolution, equation~\ref{eq:Mps}) is then $\tdis\propto \Mps$. 
Escape of stars from a more realistic, anisotropic Roche volume around the cluster is delayed relative to a spherical one \citep{2000MNRAS.318..753F}, which leads to a modified scaling of the form $\tdis\propto \Mps^x$, with $x\simeq2/3$ \citep{2001MNRAS.325.1323B}. 
For clusters without BHs, the mass-loss timescale is also $\taum\propto M^x$, such that $\dot{M} = -M/\taum \propto M^{1-x}$ \citep{2010MNRAS.409..305L}, and this scaling is often used in GC population studies  \citep{2005A&A...441..117L,pfeffer_etal18,2019MNRAS.486..331C,chen_gnedin22}.
The positive correlation between $|\dot{M}|$ and $M$ leads to a reduced mass-loss rate as the cluster approaches dissolution, that is, a gentle `skiing' $M(t)$ evolution. This allows proportionally more low-mass clusters to survive until the present, leading to GCMF with a lower $\mto$ and a broader dispersion \citep{2009MNRAS.394.2113G} than for the case of $x=1$. 

As we have shown in Section~\ref{sec:nbody}, the opposite regime of $x>1$ is needed to mimic the effect of BHs and obtain steep `jumping' $M(t)$ curves. However, with the simple scaling above this would result in a super-linear scaling between $\tdis$ and $\Mps$, which is not found in these $N$-body simulations (see Fig.~\ref{fig:tdis}).

To unify a `jumping' $M(t$) with a sub-linear dependence of $\tdis$ on the initial mass, we write the mass-loss rate generally as
\begin{equation}
\dot{M} =  \mdotref \left(\frac{M}{\Mps}\right)^{1-y} \left(\frac{\Mps}{2\times10^5\,\msun}\right)^{1-x}\frac{\omegatid}{{0.32}\,\myr^{-1}},
\label{eq:mdot}
\end{equation}
where $x>0$ is a parameter that controls the relation between $\tdis$ and $\Mps$, and $y>0$ is a parameter that controls the shape of $M(t)$. 
Here $\mdotref<0$ is the mass-loss rate at a fixed reference mass $M=2\times10^5\,\msun$ and the same reference $\omegatid$ as in equation~(\ref{eq:bm03}), that is,  for $\Vc=220\,\kms$ and $\rgeff=1\,\kpc$.

The functional form of equation~(\ref{eq:mdot}) recovers the simple constant mass-loss rate for $x=y=1$,  and the `skiing' $M(t)$ evolution with the delayed escape for $x=y\simeq2/3$. However, $x$ and $y$ do not need to be the same, and equation~(\ref{eq:mdot})  allows for both  `skiing' ($y<1$) and `jumping' ($y>1$) evolution, for any scaling between $\tdis$ and initial mass via the parameter $x$. A physical explanation for why $x$ and $y$ can be different lies in the history of disruption of clusters with the same  $M$ but different $\Mps$.
Clusters with the same $M$, but different $M/\Mps$ have different  mass fractions in BHs, which has a large effect on $\dot{M}$, as we have seen in Section~\ref{sec:nbody}. The different dependence of $\dot{M}$ on $M$ and on $\Mps$ was proposed previously by \citet{muratov_gnedin10}, based on independent arguments.

Integrating equation~(\ref{eq:mdot}) over time we find the mass evolution
\begin{equation}
M(\Mps,\omegatid,t) = \Mps\left(1 - \frac{t}{\tdis(\Mps,\omegatid)}\right)^{1/y}, 
\label{eq:Mt}
\end{equation}
for $t\le\tdis(\Mps,\omegatid)$, where
\begin{equation}
\tdis = 10\,\gyr\, \frac{2/3}{y}\, \frac{30\,\msun/\myr}{|\mdotref|}\, \frac{0.32\,\myr^{-1}}{\omegatid} \left(\frac{\Mps}{2\!\times\!10^5\,\msun}\right)^x
\label{eq:tdis}
\end{equation}
is the total lifetime, that is, the time for the cluster mass to reach zero. The mass evolution of a cluster is defined by the four parameters $x$, $y$, $\mdotref$ and $\omegatid$.

In the left panel of Fig.~\ref{fig:gcmf0} we {provide a synopsis of} the mass evolution for different choices of these parameters. The values of $x$ and $y$ also affect the shape of the GCMF, which we discuss next. 

%_____________________________
\subsection{Cluster mass function}
We define the mass function $\psi$ at time $t$ as the number of GCs in the mass range $[M, M+\dr M]$, that is, $\psi(M, \omegatid, t)= \dr N/\dr M (M, \omegatid, t)$. We can relate the mass function to the initial mass function $\psi_0(\Mps)$  as
\begin{equation}
\psi(M,\omegatid,t)= \psi_0[\Mps(M)]\left| \frac{\partial \Mps}{\partial M}(M,\omegatid,t)\right|.
  \label{eq:dndm}
\end{equation}
An analytical expression for the dependence $\Mps(M)$ (that is, the inverse of equation~(\ref{eq:Mt}) for $M(\Mps)$) can only be found for $x=y$ \citep{2009MNRAS.394.2113G}, so we cannot write the mass function analytically in a general case $x\ne y$. Instead, we find $\partial M/\partial \Mps$ from the relation given by equation~(\ref{eq:Mt}):
\begin{equation}
  \frac{\partial \ln{M}}{\partial \ln{\Mps}} =  
   1 - \frac{x}{y} + \frac{x}{y} \left(\frac{M}{\Mps}\right)^{-y}.
  \label{eq:dmdm0}
\end{equation}
By interpolating both this relation and $M(\Mps)$ from equation~(\ref{eq:Mt}), we find $\Mps(M)$ and $\partial \Mps/\partial M(M)$ to evaluate $\psi(M,\omegatid,t)$.

The right panel of Fig.~\ref{fig:gcmf0} shows cluster mass functions evolved from the initial $\psi_0(\Mps) \propto \Mps^{-2}$, for different $x$ and $y$. The logarithmic slope at low masses equals $y$, independent of $x$. This can be understood from considering the  behaviour of $\partial M/\partial \Mps$ in the limit $M \ll \Mps$. In this case $\psi(M) \approx \psi_0(\Mps) \,\frac{y}{x}\, (M/\Mps)^{y-1}$ from equations~(\ref{eq:dndm}) and (\ref{eq:dmdm0}). The logarithmic mass function $\psi(\log M)\propto M\psi(M) \propto M^{y}$.
This power-law holds for masses below the minimum initial mass of clusters still surviving at the time of observation $t$, which can be found from equation~(\ref{eq:tdis}):
\begin{equation}
  \mmin(t) = 2\times10^5\,\msun \left(\frac{y\, t\, |\mdotref|}{2\times10^5\,\msun}\right)^{1/x}.
  \label{eq:mmin}
\end{equation}

\begin{figure}
\includegraphics[width=\columnwidth]{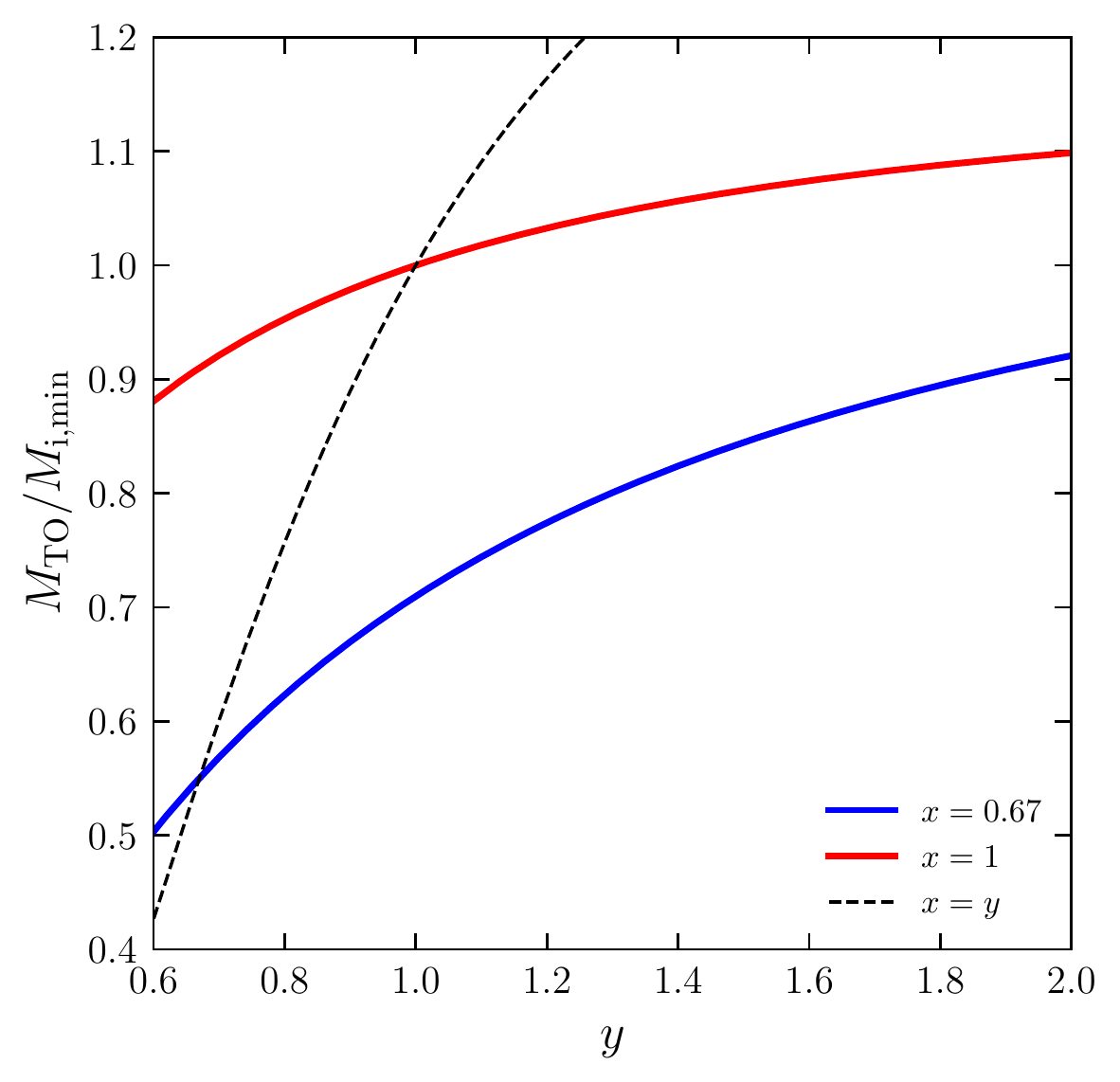}
\vspace{-5mm}
\caption{Turn-over mass $\mto$ of GCMF found numerically as a root of equation~(\ref{eq:mto}). The special case $x=y$ allows an analytic solution $\mto/\mmin = y^{1/y}$, shown by the dashed line.}
\label{fig:mto}
\end{figure}

For a given $\mdotref t$, $\mto$ increases with increasing $y$. It can be calculated by setting $d\ln{\psi}/d\ln{M} = -1$ \citep{2009MNRAS.394.2113G}. For the initial $\psi_0(\Mps) \propto \Mps^{-2}$, after some algebra this gives the equation 
\begin{equation}
  x\,(x\mu-1)\,(\mu-1)=y,
  \label{eq:mto}
\end{equation}
where $\mu \equiv (M/\Mps)^{-y}$. The root of this equation can be found numerically after we specify the relation between the initial cluster mass $\Mps$ and mass $M(t)$ after time $t$. That relation is given by equation~(\ref{eq:Mt}) and can be expressed in both masses normalized by the minimum survival mass: $(M/\mmin)^y = (\Mps/\mmin)^y - (\Mps/\mmin)^{y-x}$. The latter relation allows to convert the turnover mass $\mto$ from equation~(\ref{eq:mto}) to a ratio $\mto/\mmin$. The numerical solutions for two cases $x=1$ and $x=0.67$ are shown in Fig.~\ref{fig:mto}. They reproduce the values in Table~\ref{tab:mto} found from the evolved models.

There are three contributing factors to the higher $\mto$ when BHs are considered: (1) a larger $y$ and (2) a higher $|\mdotref|$ both increase $\mmin$ (equation~\ref{eq:mmin}) and 
 (3) a larger $y$ also results in a larger $\mto/\mmin$ (Fig.~\ref{fig:mto}). As an illustration, we highlight the difference in $\mto$ for the parameter we will use in the next section: 
 for the models with BHs ($y=1.33, \mdotref=-45\,\msun\,\myr^{-1}$), $\mto$ is a factor of $2.8 \times 1.45 \times  1.5= 6.1$ higher than for the model without BHs ($y=0.67, \mdotref=-30\,\msun\,\myr^{-1}$), for the same $x=0.67$ and the same $\omegatid$.

This simple model illustrates the general behaviour of the mass function under the new parametrization of the cluster mass loss. While many of its assumptions are not valid in the real Milky Way galaxy, the analytical expressions help us understand main effects of including BHs on the accelerated disruption. In the next section we present a more realistic model that accounts for the orbit distribution in the galaxy and matches the observations of Galactic GCs.

\begin{table}
    \centering
    \begin{tabular}{cccccc}
        \hline
        $x$ & $y$ & $\mto/\mmin$ &$ \displaystyle\log_{10}\!\mto$ & $\mu_{\log_{10}\!M}$ & $\sigma_{\log_{10}\!M}$ \\
         & & &  &[dex] \\
        \hline
       0.67 &       0.67 &       0.55 &       5.16 &       4.99 &       0.93 \\
        0.67 &       1.00 &       0.70 &       5.53 &       5.48 &       0.73 \\
        0.67 &       1.33 &       0.80 &       5.78 &       5.78 &       0.60 \\
        0.67 &       1.75 &       0.88 &       5.99 &       6.03 &       0.49 \\
        0.67 &       2.00 &       0.92 &       6.10 &       6.14 &       0.45 \\
        1.00 &       0.67 &       0.91 &       5.34 &       5.13 &       0.89 \\
        1.00 &       1.00 &       1.00 &       5.56 &       5.49 &       0.71 \\
        1.00 &       1.33 &       1.05 &       5.70 &       5.71 &       0.59 \\
        1.00 &       1.75 &       1.08 &       5.83 &       5.89 &       0.51 \\
        1.00 &       2.00 &       1.10 &       5.90 &       5.97 &       0.47 \\
          \hline
    \end{tabular}
    \caption{Turnover mass ($\mto$) and (logarithmic) mean ($\mu_{\log_{10}\!M}$) and dispersion ($\sigma_{\log_{10}\!M}$) for the mass functions shown in the right panel of Fig.~\ref{fig:gcmf0} and two additional values of $y$ ($y=1.75$ and $y=2$).}
    \label{tab:mto}
\end{table}

%%%%%%%%%%%%%%%%%%%%%%%%%%%%%%%%%%%%%%%%%%
\section{Population model}
\label{sec:model}

In this section we present a detailed model for the evolution of a GC population in a Milky Way-like galaxy. Our approach is similar to that of \citet{2001ApJ...561..751F} who start with a distribution function (DF) and then evolve the GCMF due to various disruption mechanisms which all have a dependence on the orbit. Their DF depends on isolating integrals (specific energy and angular momentum), but we express the DF directly in terms of mass, position and velocity. 
We adopt a SIS for the Galaxy, which has a potential 
\begin{equation}
\Phi(\rg) =\Vc^2\, \ln(\rg),
\label{eq:phiR}
\end{equation}
and we assume the circular velocity of $\Vc=220\,\kms$.

%______________________
\subsection{Initial conditions}
\label{ssec:ics}

We write the initial DF, that is, the phase-space density of clusters in the galaxy, as a function of $\Mps$, galactic position ($\rgvec$) and galactic velocity ($\vgvec$)
\begin{equation}
f_0(\Mps,\rgvec,\vgvec) = \psi_0(\Mps)\, n_0(\rg)\, F(R,\Vr, \Vt).
\label{eq:f0}
\end{equation} 
Here $\psi_0(\Mps)$ describes the initial GCMF, $n_0(\rg)$ the radial number density profile, and $F(R, \Vr, \Vt)$ the radius-dependent velocity distribution. Because we will consider the effect of radially-biased velocity anisotropy for the GC orbits, we define the velocity distribution in terms of the radial velocity ($\Vr$) and the tangential velocity ($\Vt$).
The phase-space density is normalised such that $\int_0^{\infty}\dr^3\vgvec\int_{\Rlo}^{\Rup}\dr^3\rgvec\int_{\Mlo}^{\Mup}\dr \Mps\, f_0(\Mps,\rgvec,\vgvec)=1$, where we adopt the following boundary values: $\Rlo=1\,\kpc, \Rup=100\,\kpc, \Mlo=10^4\,\msun$, and $\Mup=10^8\,\msun$. We discuss the effect of varying $\Mlo$ in Section~\ref{sec:discussion} and  next we discuss the functional forms for each contribution to $f_0$. 

For $\psi_0(\Mps)$ we adopt a power-law with an exponential truncation
\begin{equation}
\psi_0(\Mps) \propto \Mps^{-\alphaM}\exp\left(-\frac{\Mps}{\Ms}\right),
\label{eq:icmf}
\end{equation}
where $\Ms$ is the truncation mass, which we take either $\Ms\rightarrow\infty$ (the `power law' model considered in the previous section), or $\Ms\simeq10^6\,\msun$ (the `Schechter' model). We fix the power-law index $\alphaM=2$ in all models, as it is a common value found for young massive clusters in nearby galaxies. Although it cannot be ruled out that GCs formed with a more peaked initial GCMF, here we adopt the hypothesis that massive star clusters form with a universal GCMF at all redshifts. The constant of proportionality is found from the requirement that $\int_{\Mlo}^{\Mup}\psi_0\, \dr \Mps = 1$.

The velocity distribution is assumed to be Gaussian in all three components, such that
\begin{equation}
F(R,\Vr, \Vt) = \frac{\exp\left[-\Vr^2/(2\sigr^2(R))\right]}{\sqrt{2\pi}\sigr(R)}\frac{\exp\left[-\Vt^2/\sigt^2(R)\right]}{\pi\sigt^2(R)}.
\end{equation}
Here $\sigt(R)=\langle\Vt^2(R)\rangle^{1/2}$ is the root-mean-square tangential velocity at radius $R$ and $\sigr(R)=\langle \Vr^2(R)\rangle^{1/2}$ is the root-mean-square radial velocity at radius $R$. For an isotropic velocity distribution $\sigt^2=2\sigr^2$. It satisfies $\int\dr^3 \vgvec F(R,\Vr, \Vt)=2\pi \int_0^{\infty}\Vt \dr \Vt\int_{-\infty}^{\infty}\dr \Vr\, F(R,\Vr, \Vt)=1$. 
We adopt an anisotropy profile of the GC system of the form
\begin{align}
\beta(\rg) &\equiv 1-\frac{\sigt^2(\rg)}{2 \sigr^2(\rg)} \nonumber\\
  &= \frac{1}{1+(\rga/\rg)^\delta},
  \label{eq:beta}
\end{align}
{where $\delta>0$.}
This profile results in isotropy for $\rg\ll\rga$  and radial orbits for $\rg\gg \rga$, with the parameter $\delta$ determining how quickly $\beta(\rg)$ rises. DFs that include radial anisotropy with an $\exp(-J^2)$ term, where $J$ is the specific angular momentum, result in a $\beta(\rg)$ profile as in equation~(\ref{eq:beta}) with $\delta=2$ \citep{1915MNRAS..75..366E,1963MNRAS.125..127M,1979PAZh....5...77O,1985AJ.....90.1027M}. The present-day anisotropy profile of Milky Way GCs as derived from line-of-sight velocities  and {\it Gaia} proper motions \citep{2019MNRAS.484.2832V}  is better described by $\delta\simeq1$. Here we will vary $\delta$, together with $\rga$, to match the $\beta$ profile of the observed clusters {(see the description of Model (3) in Section~\ref{ssec:modelpars})}. 

To find $\sigt(R)$ and $\sigt(R)$ we need to define $n_0(\rg)$ and solve the radially-anisotropic Jeans equation \citep[eq. 4.215 in ][]{2008gady.book.....B} with GCs as tracer particles in the Galactic potential (equation~\ref{eq:phiR}). A convenient choice  for $n_0(\rg)$ is
\begin{equation}
 n_0(\rg) \propto \frac{\rg^{-\alphaR}}{\left[1+(\rg/\rga)^\delta\right]^{2/\delta}},
 \label{eq:n0}
\end{equation}
because it results in a constant radial dispersion $\sigr=\Vc/\alphaR^{1/2}$.
For $\rg\ll\rga$, the profile is $n_0(\rg)\propto \rg^{-\alphaR}$, and for $\rg\gg\rga$ it is $n_0(\rg)\propto \rg^{-\alphaR-2}$. Fully isotropic models ($\rga\rightarrow\infty$) have a single power-law $n_0(\rg)\propto \rg^{-\alphaR}$. The constant of proportionality  is found from the requirement that $\int n_0(R)\,\dr^3\rgvec =4\pi\int_{\Rlo}^{\Rup}n_0(R)R^2\dr R = 1$.
Combined with the expression for $\beta(R)$, we find that $\sigt^2(R) = 2\sigr^2[1+(\rg/\rga)^\delta]^{-1}$. We now have a fully analytic form for $f_0(\Mps,\rgvec,\vgvec)$ (equation~\ref{eq:f0}) and describe next how we evolve it to the present age of GCs. 

%____________________________________
\subsection{Evolving the GC population}
To evolve the mass function as a function of $R$, we need to obtain the effective tidal field strength $\omegatid=\Vc[\Rp(1+\ecc)]^{-1}$ from $\Vr, \Vt$ and $\rg$. For the case of a SIS, $\Rp(\rg, \Vr, \Vt)$ and $\Ra(\rg, \Vr, \Vt)$ are the radii where $\Vr=0$, which are roots that need to solved numerically from the orbital energy and angular momentum \cite[see, for example, Section~2.1 of ][]{1999ApJ...515...50V}, which then provides $\ecc=(\Ra-\Rp)/(\Ra+\Rp)$. Because of the scale-free nature of the SIS, we do this once and use interpolation to find $\omegatid(\rg, \Vr, \Vt)$. 

We first find the present-day phase-space density $f(M,\rgvec, \vgvec,t) = \psi(M,\omegatid,t)\, n_0(R)\, F(R,\Vr, \Vt)$, where $\psi(M,\omegatid,t)$ is given by equation~(\ref{eq:dndm}) and we recall that $\omegatid=\omegatid(R,\Vr,\Vt)$.
We then integrate over all velocities to obtain the mass function as a function of $\rg$
\begin{align}
\psi(M,\rg,t) &= \int  f(M, \rgvec,\vgvec,t)\, \dr^3\vgvec,\nonumber\\
&= 4\pi R^2n_0(R)\!\int \dr^3\vgvec  \psi(M,\omegatid,t)F(R,\Vr, \Vt).
\label{eq:fmrt}
\end{align}
The fraction of surviving clusters is then given by $\fsurv = \int_{\Mlo}^{\Mup} \dr M \int_{\Rlo}^{\Rup}\dr R\, \psi(M, R, t)$. To compare the model to the observations we multiply $\psi(M,\rg,t)$ by $\NGC/\fsurv$, with $\NGC=156$ being the total number of GCs in the Milky Way for which a luminosity and $\rg$ are available \citep{1996AJ....112.1487H,2010arXiv1012.3224H}. We summarise the various definitions of the DF and the mass function in Table~\ref{tab:popvar}.

 \begin{table}
    \centering
    \begin{tabular}{ll}
        \hline
Description & Parameter  \\\hline
Galaxy: & $\Vc=220\,\kms$    \\ 
Mass loss: & $\mdotref=-30~{\rm or}\,-45\,\msun\,\myr^{-1}$  \\
               & $x=2/3$  \\
              & $y=2/3$ or $4/3$  \\
Initial GCMF: & $\alphaM=2$  \\
           & $\Ms=10^6\,\msun\,{\rm or}\,\infty$  \\
$n_0(\rg)$: & $\alphaR=3.5$ or $4.5$  \\
$\beta(\rg)$ and $n_0(\rg)$:      & $\rga=5\,\kpc$ or $\infty$  \\
      & $\delta=0.5$  \\
         \hline
    \end{tabular}
    \caption{Overview of the nine parameters of the GC population model.}
    \label{tab:params}
\end{table}

 \begin{table}
\def\arraystretch{2.5}%  1 is the default, change whatever you need
    \centering
    \begin{tabularx}{\columnwidth}{lll}
        \hline
Function & Definition & Description  \\\hline
$f_0(\Mps, \rgvec,\vgvec)$ & $\displaystyle\frac{\dr^7\!N}{\dr \Mps\,\dr^3\!\rgvec\, \dr^3\!\vgvec}$ & Initial DF (after stellar evolution)\\
$f(M, \rgvec,\vgvec, t)$ & $\displaystyle\frac{\dr^7 N}{\dr M\,\dr^3\!\rgvec\, \dr^3\!\vgvec}$ &  DF (after evaporation)\\
$\psi_0(\Mps)$ & $\displaystyle\frac{\dr N}{\dr \Mps}$ & Initial GCMF (after stellar evolution)\\
$\psi(M, \omegatid, t)$ & $\displaystyle\frac{\dr N}{\dr M}$ & Evolved GCMF  for a single $\omegatid$\\
$\psi(M, R, t)$ & $\displaystyle\frac{\dr^2\!N}{\dr M\dr\rg}$ & Evolved GCMF at $\rg$ (equation~\ref{eq:fmrt})\\
$\psi(M, t)$ & $\displaystyle\frac{\dr N}{\dr M}$ &   $\psi(M, R, t)$ integrated over  $\rg$\\
         \hline
    \end{tabularx}
    \caption{Overview of the functions used in the population model.}
    \label{tab:popvar}
\end{table}

%__________________________________________________-
\subsection{Model parameters}
\label{ssec:modelpars}
In Table~\ref{tab:params} we summarise the nine parameters of the GC population model we described above, including the adopted values. The parameters with a single mentioned value are fixed in all models. For some parameters we adopt two values, in order to study their effect on the resulting GCMF. We use these parameters to solve eight models summarised in Table~\ref{tab:models} and described below.

 \begin{table*}
    \centering
    \begin{tabular}{llcccccccccc}
        \hline
Model& \multicolumn{5}{c}{Description} & \multicolumn{5}{c}{Parameters}\\\hline
     & Name                 & Anisotropy & ICMF      & $\feh$ gradient & Past tidal & $\rga$   & $\delta$ & $\alphaR$ & $\Ms$         & $\mdotref$ & $y$\\
     &                      &            &           &                 &            & [kpc]    &           &          & [$\msun$]     &[$\msun/\myr$]  \\ \hline
(1)   & no BHs               & no         & power law & no              & no         & $\infty$ &  $-$      & $4.5$    & $\infty$      & -30 & $0.67$\\  
(2)  & BHs                  & no         & power law & no              & no         & $\infty$ &  $-$      & $4.5$    & $\infty$      & -45 & $1.33$\\  
(3)  & BHs + A (Anisotropy)      & yes        & power law & no              & no         & $5$      &  $0.5$      & $3.5$    & $\infty$      & -45 & $1.33$\\ 
(4)  & BHs + S (Schechter)       & no         & Schechter & no              & no         & $\infty$ &  $-$      & $4.5$    & $10^6$ & -45 & $1.33$\\
(5)  & BHs + F ([Fe/H] gradient) & no         & power law & yes             & no         & $\infty$ &  $-$      & $4.5$    & $\infty$      & -45 & $1.33$ \\ 
(6)  & BHs + P (Past evolution)  & no         & power law & no              & yes        & $\infty$ &  $-$      & $4.5$    & $\infty$      & -45 & $1.33$ \\  
(7) & BHs + A+S        & yes        & Schechter & no             & no        & $5$      & $0.5$      &  $3.5$    & $10^6$ & -45& $1.33$\\ 
(8) & BHs + A+S+F+P        & yes        & Schechter & yes             & yes        & $5$      & $0.5$      &  $3.5$    & $10^6$ & -45 & $1.33$\\
        \hline
    \end{tabular}
    \caption{Different population models shown in Figs.~\ref{fig:gcmf_problem1} and \ref{fig:gcmf_problem2}. In all models $x=0.67$, $\alphaM=2$, $\Vc=220\,\kms$.}
    \label{tab:models}
\end{table*}

\begin{figure}
\includegraphics[width=\columnwidth]{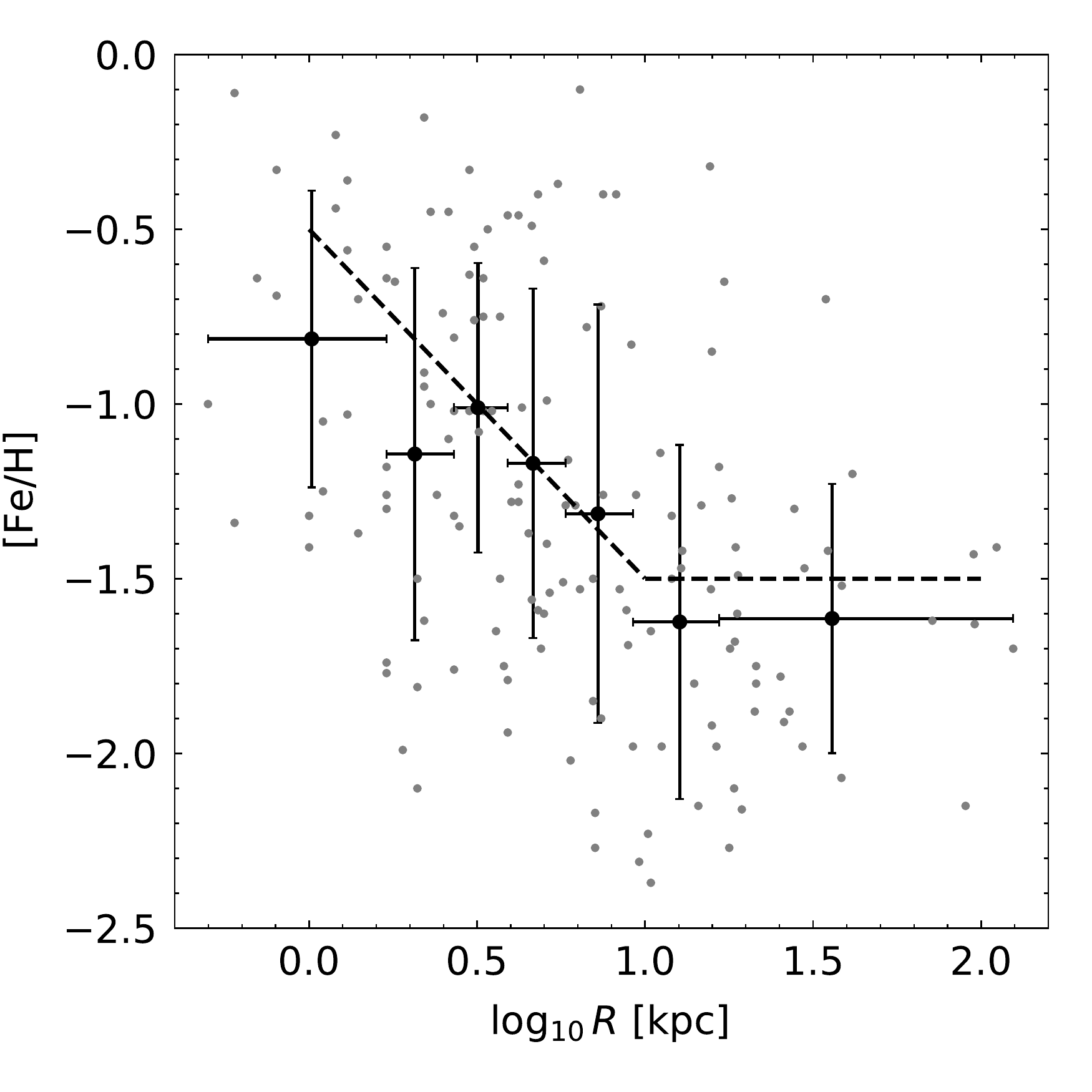}
\vspace{-7mm}
\caption{Metallicities of Milky Way GCs, as individual points (gray symbols) and in radial bins. Dashed line shows a simple analytic approximation, which for $\rg<10\,\kpc$ has a logarithmic slope of $-2/3$: $\feh(\rg)=-1.5+(1-\log_{10}\rg)/1.5$.}
\label{fig:fehrg}
\end{figure}

\begin{enumerate}
\item {\bf Model (1), no BHs:} This model serves as a starting point and defines the magnitude of the GCMF problem by considering the simplest case: an isotropic velocity distribution $\beta(\rg)=0$. Integrating over all orbits, we find that the average mass-loss rate at $\rg$ is a factor of $\sim3.1$ higher than that of the circular orbit at that $\rg$.
The initial GCMF is a power law and for the mass loss parameters we adopt the values found in models without BHs (equation~\ref{eq:bm03}), that is, $\mdotref=-30\,\msun\,\myr^{-1}$ and $x=y=2/3$. This model roughly describes the contribution of evaporation in the hierarchical models of \citet{pfeffer_etal18} and \citet{choksi_gnedin19b} at low redshift. 
\item {\bf Model (2), BHs:} Here we study the effect of higher $|\dot{M}|$ and $y$ due to BHs. We base the values on the $N$-body models with $\rhoratio_0\simeq30$ from Section~\ref{sec:nbody}, which can be described by $\mdotref=-45\,\msun\,\myr^{-1}$ and $y=4/3$.
In these models the effect of BHs is moderate compared to the two sets of $N$-body models with lower $\rhoration$ shown in Fig.~\ref{fig:mdot_nbody}, which have higher $|\mdotref|$ and $y$.
We assume the same $\rhoration\simeq30$ and  $\rhn/\rjeff\simeq0.05$ (Table~\ref{tab:rho}) for all clusters. This is of course not realistic, because real clusters have a spread in these parameters and the orbits evolve in time, but it serves as an approximation for the average filling factor of clusters. We discuss this point more in Section~\ref{ssec:rhohn}.

In the next four models, we add a single physical effect to the BHs, which each reduce the gradient of $\mto(\rg)$:

\item {\bf Model (3), BHs + A (Anisotropy):} Here we add radially-biased anisotropy by choosing $\rga=5\,\kpc$  with a relatively slowly rising $\beta(\rg)$ ($\delta=0.5$). These values were chosen such that in the final Model (8) the anisotropy profile of the surviving clusters is similar to the observed profile. Anisotropy increases $|\dot{M}|$ at large $\rg$, thereby reducing the gradient of $\mto(\rg)$.

\item {\bf Model (4), BHs + S (Schechter):} Here we add a Schechter truncation mass of $\Ms=10^6\,\msun$ as found by \citet{2007ApJS..171..101J} from fits of `evolved Schechter functions' of the Milky Way GCMF. If the amount of mass lost is comparable to $\Ms$, the turnover mass only increases slowly for any additional mass loss \citep{2007ApJS..171..101J, 2009MNRAS.394.2113G}, so this truncation mass reduces the gradient of $\mto(\rg)$ in the inner galaxy. 

\item {\bf Model (5), BHs + F (\feh\ gradient):} Here we consider the effect of the GC metallicity gradient in the galaxy. In Section~\ref{ssec:metallicity} we showed that more metal-rich clusters ($\feh\gtrsim-0.5$), with the same initial density ($\rhoration\simeq30$) eject all BHs and evolve similarly to clusters without BHs. 
This means that in the inner galaxy, $|\dot{M}|$ is lower than in Model (2). Fig.~\ref{fig:fehrg} shows that in the range $-0.5\lesssim \log_{10}(\rg/\kpc)\lesssim1$, $\feh$ of Milky Way GCs decreases from $-0.5$ to $-1.5$. 
We mimic the effect of a \feh-gradient by adopting $\rg$-dependent relations for $\dot{M}$ and $y$ for $\rg\le10\,\kpc$:
\begin{align}
\mdotref(\rg) &= -30\,\msun\,\myr^{-1}\, \left(1+\frac{1}{2}\log_{10}(\rg) \right) \\
y(\rg) &=\frac{2}{3}+\frac{2}{3}\log_{10}(\rg).
\label{eq:mdotry}
\end{align}
At $\rg=10\,\kpc$ these relations result in the same values as Model~(2) ($\mdotref=-45\,\msun\,\myr^{-1}$, $y=4/3$), and at $\rg=1\,\kpc$ they give the values found for clusters without BHs ($\mdotref=-30\,\msun\,\myr^{-1}$, $y=2/3$, equation~\ref{eq:bm03}), thereby reducing the effect of BHs on $\dot{M}$ and therefore the gradient of $\mto(\rg)$.
\item {\bf Model (6), BHs + P (Past evolution):} Here we include an approximate correction for the past tidal evolution of clusters from a full hierarchical model, described in Appendix~\ref{sec:hierarchical}. As a result we multiply $\dot{M}$ by $(\rgeff/4)^{1/2}$ at $\rgeff>4$~kpc.
\item {\bf Model (7), BHs + A+S:} In this model we combine the effect of anisotropy and the Schechter cutoff mass. 
\item {\bf Model (8), BHs + A+S+F+P:} Here we include all effects described in Models (2)-(6). This model represents a realistic way of modelling cluster evolution. 
\end{enumerate}

%%%%%%%%%%%%%%%%%%%%%%%%%%%%%%%%%%%%%%%%%%%%%%%%%
\section{Results of population models}
\label{sec:results}

\begin{figure*}
\includegraphics[width=0.99\textwidth]{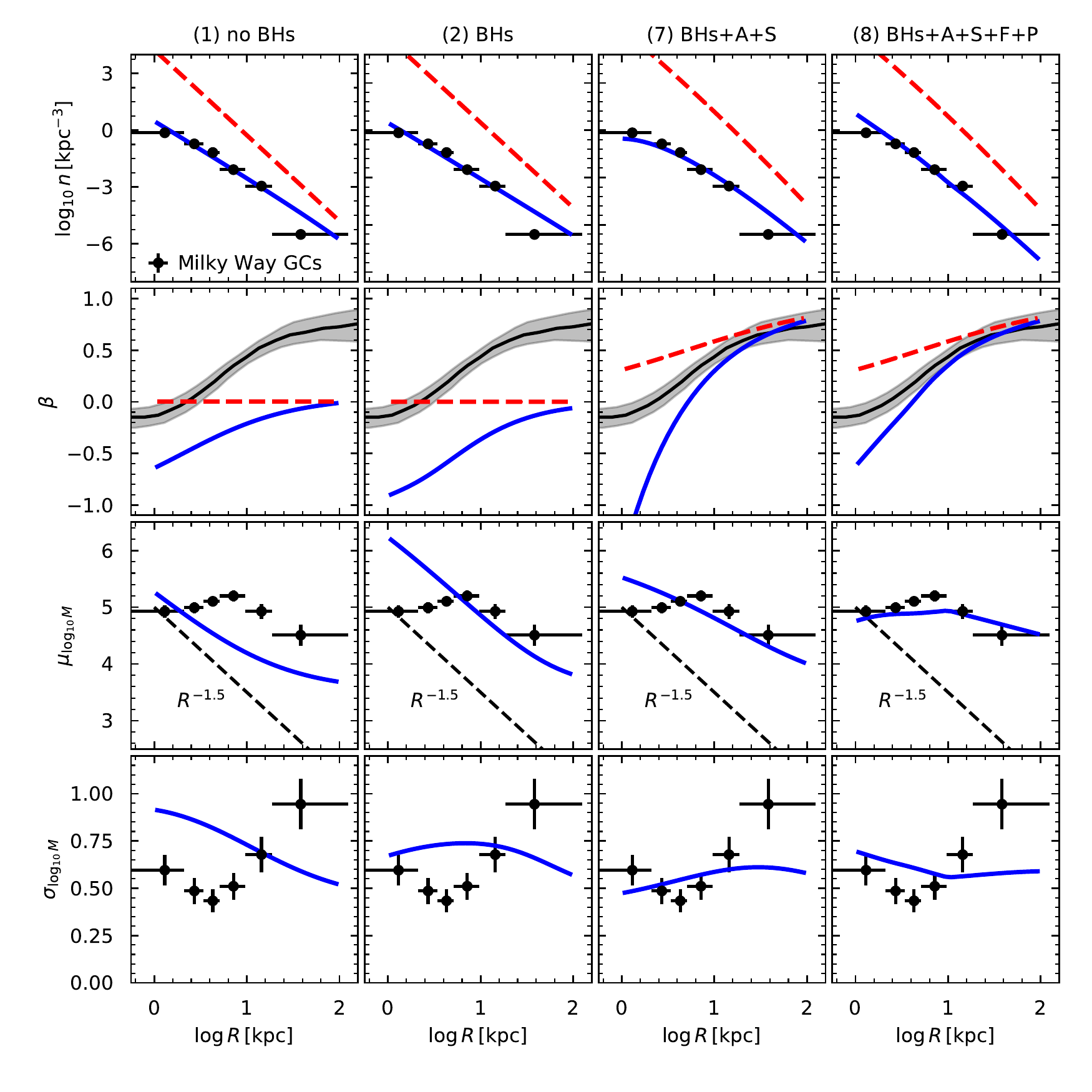}
\vspace{-7mm}
\caption{Different ingredients in the GCMF model compared. \textbf{A} stands for Anisotropy, \textbf{S} for Schechter, \textbf{F} for \feh\ gradient, \textbf{P} for past evolution. Red dashed lines in the top panels are the initial distributions, solid blue lines are the final distributions, black data points with error bars {and black solid line with grey shaded regions} are observations.
}
\label{fig:gcmf_problem1}
\end{figure*}

\begin{figure*}
\includegraphics[width=0.99\textwidth]{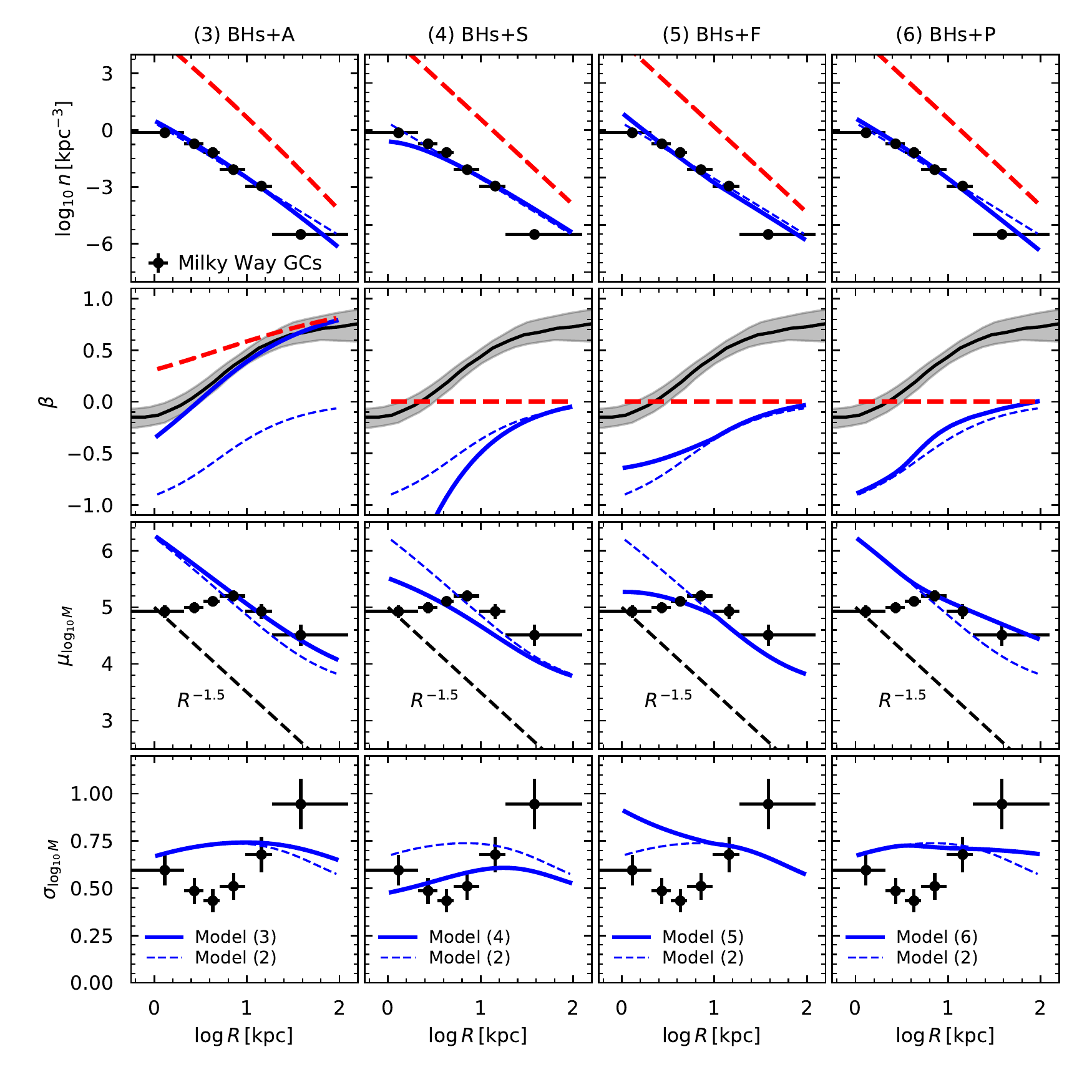}
\vspace{-7mm}
\caption{{As Fig.~\ref{fig:gcmf_problem1}, but now for Models (3-6), showing in each column the effect BHs and one additional ingredient (\textbf{A} = Anisotropy; \textbf{S} = Schechter; \textbf{F} = \feh\ gradient; \textbf{P} = past evolution).  }}
\label{fig:gcmf_problem2}
\end{figure*}

In this section we discuss the results of the eight models summarised in Table~\ref{tab:models}. We compare the models to 156 GCs with luminosities and $\rg$ in the Harris catalogue \citep{1996AJ....112.1487H,2010arXiv1012.3224H}. We adopt a mass-to-light ratio $M/L_V = 1.8$ from \citet{2020PASA...37...46B} and quantify the shape of the GCMF at different radii by $\mulog \equiv\langle\log_{10}(M/\msun)\rangle$ and the dispersion of the logarithmic mass distribution, $\siglog$, {for $M\ge10^3\,\msun$}. The mean of the logarithm of mass is a reasonable approximation to $\mto$ (see Table~\ref{tab:mto}).
For the anisotropy we {consider clusters with $M>10^5\,\msun$ and compare to} the  results from {\it Gaia} DR2 by \citet{2019MNRAS.484.2832V}. All models are displayed in Figs.~\ref{fig:gcmf_problem1} and \ref{fig:gcmf_problem2}.

\subsection{Global properties}
\begin{enumerate}
\item {\bf Model (1), no BHs:} This model quantifies the magnitude of the GCMF problem: $\mulog$ reproduces the data only at $\rg\simeq1\,\kpc$ and then it declines as $\rg^{-1/x}=\rg^{-3/2}$, which is a known scaling for $\mto(\rg)$ for an initial GCMF that is a power-law with logarithmic slope of $-2$ and $\tdis\propto M_0^x$ \citep{2009MNRAS.394.2113G}. At  $\rg\simeq100\,\kpc$, $\mulog$ is a factor of $\sim10$ too low. The model also underestimates $\beta$ at all radii, because preferentially radial orbits are removed at small radii while the initial $\beta$ is too low at large $\rg$ {and evolves very little}. The power-law initial GCMF and low value for $x$ also result in a very wide GCMF ($\slogm\simeq0.75$) compared to the observed width ($\slogm\simeq0.5$). The model width is also decreasing with $\rg$, while the data show an increase. From this it is clear that evaporation  of clusters without BHs is not able to explain the shape of the GCMF. Adding radial anisotropy within the constraints of the observed $\beta(\rg)$ increases $\mulog$ by only $\sim0.3\,$dex at large $\rg$ (not shown), and is therefore not sufficient.
\item {\bf Model (2), BHs:} Here we only change $\mdotref$ and $y$ to mimic the effect of BHs on $\dot{M}$. This model almost reproduces $\mulog$ at $\rg\gtrsim10\,\kpc$, suggesting that the effect of BHs  alleviates a large part of the GCMF problem. This model gives rise to the same scaling  $\mulog\propto \rg^{-3/2}$, so when adding BHs the problem is that $\mto$ is too high in the inner galaxy ($\lesssim10\,\kpc$).

\begin{figure*}
\includegraphics[width=0.98\textwidth]{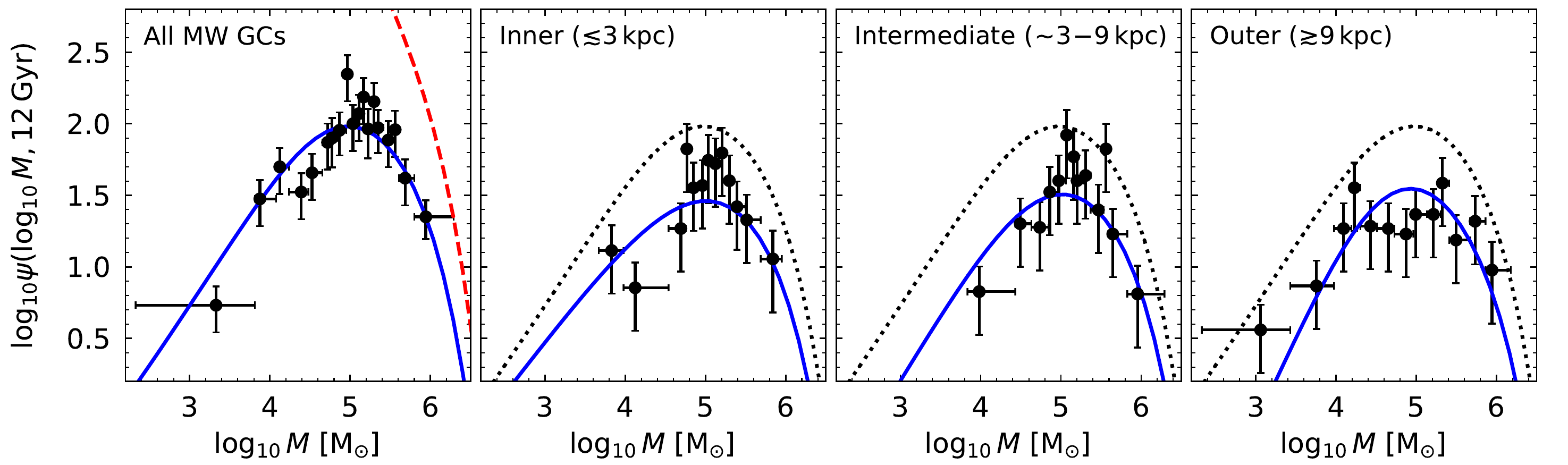}
\vspace{-1mm}
\caption{Comparison between the GCMF of Model (8) and the observed MF of Galactic GCs from the Harris catalogue for all 156 GCs (left) and at three Galactocentric radius bins containing equal number of GCs (52) each. The dashed (red) line shows the initial GCMFs. The dotted (black) lines in the three right panels show the total GCMF model for reference.}
\label{fig:gcmfall}
\end{figure*}

Models (3)--(6) present four additional ingredients that all reduce the (logarithmic) slope of the $\mulog(\rg)$ relation. The individual effects can be seen in Fig.~\ref{fig:gcmf_problem2}.
\item {\bf Model (3), BHs + A (Anisotropy):} In this model we add radial anisotropy to the effect of BHs. We find that for $\rga=5\,\kpc$ and $\delta=0.5$ the observed $\beta(\rg)$ profile is well reproduced. Compared to Model (2), radially-biased anisotropy increases $\mulog$ by $\sim0.5\,$dex at $\rg\gtrsim10\,\kpc$, improving agreement with the observations at those large radii.
\item {\bf Model (4), BHs + S (Schechter):} The addition of the  exponential truncation in the initial GCMF 
reduces $\mulog$ at $\rg\simeq1\,\kpc$ by nearly an order of magnitude. It also narrows the GCMF to approximately the correct width $\slogm\simeq0.5$.
\item {\bf Model (5), BHs + F (\feh\ gradient):} The inclusion of the metallicity gradient 
reduces $\dot{M}$ mainly at $\rg\lesssim3\,\kpc$, decreasing $\mulog$ and increasing $\slogm$ there.
\item {\bf Model (6), BHs + P (Past evolution):} The past tidal evolution reduces the gradient of $\mulog(\rg)$ from $\rg^{-3/2}$ to approximately $\rg^{-0.9}$.
\item {\bf Model (7), BHs + A+S:} The combined effect of anisotropy and a Schechter cutoff reduces the decline of {$\mulog$}, but still the disruption at small(large) $\rg$ is overestimated(underestimated) slightly.
\item {\bf Model (8), BHs + A+S+F+P:} This model combines all four  effects (A+S+F+P) in addition to the BHs. We note that because of the relatively small $\delta=0.5$, the logarithmic slope of the initial density profile is steeper than $-3.5$ at all radii: it reduces from $-4.1$ at $1\,\kpc$ to $-5.1$ at $100\,\kpc$, and at $\rga=5\,\kpc$ the slope is $-4.5$. 
In this model the gradient of {$\mulog$} is substantially reduced compared to the power law Model (2) and closely follows the observed $\mulog(\rg)$. 
This model reproduces well the number density profile, the anisotropy profile, and the shape of the GCMF at different Galactic radii. 
The model orbits are a bit too tangentially biased ($\beta<0$) at $\rg\lesssim3\,\kpc$. We interpret this as an artefact of our assumption of a static potential in which orbits do not isotropise due to interactions with the galactic bar, other GCs and infalling satellite galaxies. We expect that these effects in a real galaxy would lead to an isotropic velocity distribution of inner GCs.
\end{enumerate}

%________________________________________
\subsection{Mass function}
In Figs.~\ref{fig:gcmf_problem1} and \ref{fig:gcmf_problem2} we used the (logarithmic) mean and dispersion of the GCMF as measures of the GCMF shape. It is also instructive to look at the full GCMF in different Galactocentric radius bins. The GCMF for all GCs is obtained from $\psi(M,t) = \int_{\Rlo}^{\Rup}  \psi(M,R,t)\dr \rg$, with $\psi(M, R, t)$ from equation~(\ref{eq:fmrt}). The integration boundaries can be changed to obtain the GCMF in different radial intervals. Fig.~\ref{fig:gcmfall} shows the total mass function for Model (8) and in three radial bins. For the observational data we used radial bins with equal number of GCs (52) and the histograms were constructed with equal number of GCs: eight in the total sample and four in the three $\rg$-bins. For the model we also defined the radial bins to contain exactly 1/3 of the total number of GCs, which results in slightly different bin edges {than for the observations} because the number density profile {of the model} is slightly steeper. We do this because we are mostly interested in comparing the shapes, rather then the vertical scaling. The model GCMF shows good resemblance to the observed one, reproducing the slightly lower $\mto$ and larger width at high $\rg$.

%%%%%%%%%%%%%%%%%%%%%%%%%%%%%%%%%%%%%%%%%%%%%%%%%%
\section{Discussion}
\label{sec:discussion}
%________________________
\subsection{Contribution to field stars}
The vast majority of GCs in our model do not survive, so an important check is to compare the contribution of dissolved clusters to the field stars. 
The total mass lost from star clusters with $\Mps>10^4\,\msun$ in Model (8) is $\Delta M \simeq 5\times10^8\,\msun$ and here we discuss the implications.
The total mass of the Galactic halo is $1.4\times10^9\,\msun$ \citep{2019MNRAS.490.3426D}, so if all lost mass ended up in the halo, then roughly one third of the stellar halo would be made out of disrupted star clusters. If we adopt a lower limit of $10^2\,\msun$ this fraction approximately doubles. This appears in tension with the results of \citet*{2015MNRAS.448L..77D} who find that the ratio of the number of blue stragglers over blue horizontal branch  stars in the halo is more similar to that in dwarf galaxies than in surviving GCs. It must be noted that this ratio in low-mass clusters is closer to what is found in the field than in massive clusters \citep{2015MNRAS.448L..77D}.
Also, the contribution of stars from GCs to the halo is more important in the inner Galaxy where tides are stronger. 
To quantify this, we determined the radial density profile of mass lost from GCs, $\rho_\star(\rg)$.
We find $\rho_\star(\rg)$ by subtracting the mass in surviving clusters from the initial mass of the GC population as a function of $\rg$ 
\begin{align}
\rho_\star(\rg)&= \frac{\NGC}{\fsurv}\left[\int_{\Mlo}^{\Mup}\Mps\psi_0(\Mps)n_0(\rg) \,\dr \Mps - \right.\nonumber\\
&\hspace{2cm} \left.\frac{1}{4\pi\rg^2}\int_{\Mlo}^{\Mup}M\psi(M, \rg,12\,\gyr)\, \dr M\right].
\label{eq:DeltaM}
\end{align}
Here $\psi(M,R,t)$ is the present-day GCMF (equation~\ref{eq:fmrt}) and $\psi_0(\Mps)$ and $n_0(\rg)$ are the initial GCMF and initial number density profile from equation~(\ref{eq:f0}). 
Note that this expression is only approximate, because it assumes that the mass is lost at $\rg$ while in reality the escaped stars  follow a distribution between the pericentre and apocentre distance of the orbit. Nevertheless, equation~(\ref{eq:DeltaM}) provides a useful estimate that can be compared to observational data. 

In Fig.~\ref{fig:rhohalo} we compare $\rho_\star(\rg)$ to the stellar halo density profile from APOGEE \citep{2021MNRAS.500.5462H}.
Because the APOGEE data only include stars with $-2.5<\feh<-1$, we multiply our model predictions by a correction factor $\fmp(\rg)\le1$ that approximates the fraction of GCs with $\feh<-1$ as a function of $\rg$. From the Harris catalogue we find that this fraction for GCs today is well described by $\fmp=\left[1.5+\log_{10}(\rg/\kpc)\right]/3.5$ for $1<\rg/\kpc<100$. It increases from $\fmp(1\,\kpc)\simeq0.4$ to $\fmp(100\,\kpc)=1$. At the smallest $\rg$ in the APOGEE data ($1.5\,\kpc$), the contribution of mass lost from GCs to the total halo mass is nearly $70\%$. 
This seems extreme, but we note that at $15\,\kpc$ the fraction drops to $\sim10\%$, so there is no tension with the conclusion of \citet{2015MNRAS.448L..77D}, because their sample considered stars at $\rg\gtrsim10\,\kpc$. Also, we note that GC may form in a disc-like configuration \citep{2005ApJ...623..650K, meng_gnedin21} and {later} scatter into the halo. The fast disrupting low-mass clusters therefore may contribute more to the thick disc and/or the bulge, which are $5-10$ times more massive than the halo, respectively.

%________________________
\subsection{Nitrogen-rich stars}
\begin{figure}
\includegraphics[width=\columnwidth]{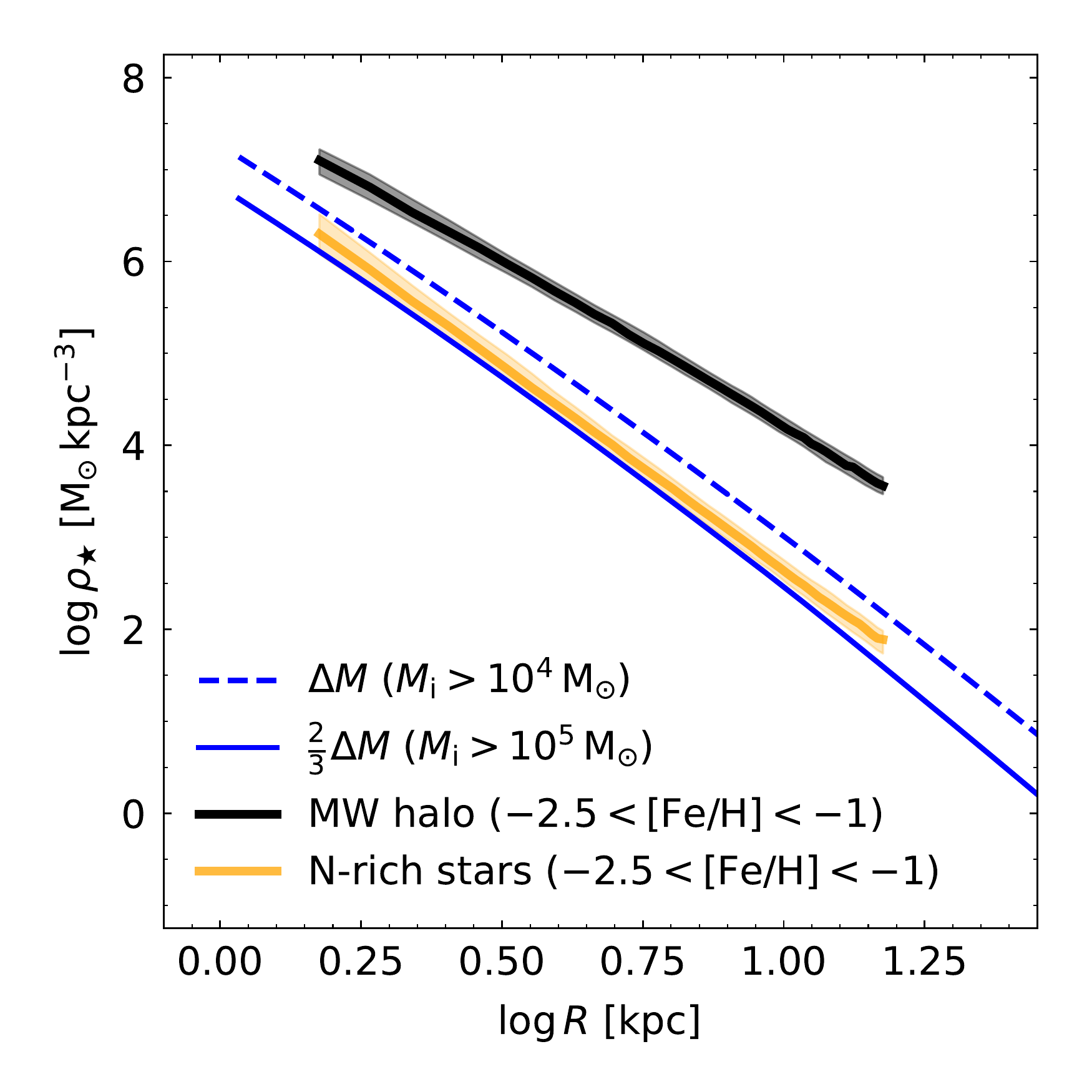}
\vspace{-8mm}
\caption{Comparison between the density profiles of stellar mass stripped from GCs ($\Delta M$) and the N-rich stars in the Galactic halo \citep{2021MNRAS.500.5462H}, which are expected to originate from massive GCs. We compare this to $\Delta M$ from GCs with initial masses $\Mps> 10^5\,\msun$ and assume that {2/3} of their stars are N rich. The APOGEE data do not include stars with $\feh \ge -1$, so we multiply our model prediction (grey dashed lines) by an approximation for the fraction of metal-poor GCs as a function of radius $f(\rg) = (1.5+\log(\rg))/3.5$, for $\rg\le100\,$kpc, which describes the present-day fraction of Milky Way GCs with $\feh<-1$.}
\label{fig:rhohalo}
\end{figure}

Most mass is lost from now-disrupted low-mass clusters, and  it is challenging to identify these stars as having originated from a cluster, because their streams will have phase mixed long ago. However, massive clusters ($\gtrsim10^5\,\msun$) have anomalous light-element abundances, manifesting as anti-correlations in N-C, Na-O and sometimes Al-Mg \citep{2018ARA&A..56...83B}, and these chemical imprints are preserved when stars are lost from the cluster. 
Stars with such abundances have also been found in the (inner) halo \citep[][]{2016ApJ...825..146M,2017MNRAS.465..501S,2021MNRAS.500.5462H,2022MNRAS.514..689B}. As another test, we compare $\rho_\star(\rg)$ of stars originating from massive clusters to the density profile of N-rich stars found in APOGEE by \citet{2021MNRAS.500.5462H}. We assume that 2/3 of GC stars with $\Mps\ge10^5\,\msun$ ($\ge \fsev10^5\,\msun$ after stellar mass loss) have anomalous abundances. In Fig.~\ref{fig:rhohalo} we show that the predicted $\rho_\star(\rg)$ of N-rich stars matches the observed profile from APOGEE very well. This suggests that these N-rich halo stars have a GC origin. A follow-up test is to look for clustering of stars in energy and angular momentum space (or action-angles), because a GC origin predicts that the N-rich stars are more clustered than the rest of the halo stars, as they originated from more massive clusters, which disrupted more recently. Additionally, at a given $\rg$, most mass is lost from the GCs that have the most radial orbits, so we predict that the N-rich stars are preferentially on radial orbit ($\beta \sim 0.5$). {Indeed, N-rich stars are on highly eccentric orbits \citep{2019MNRAS.488.2864F}, but  not more eccentric than normal metal-poor stars \citep{2020ApJ...891...28T}.  }

%________________________
\subsection{Specific frequency as a function of metallicity}
Another aspect to consider is GC disruption as a function of $\feh$.
In our model, metal-poor GCs have a higher $|\dot{M}|$ than metal-rich GCs for the same $\omegatid$. If we assume that all star formation happens in clusters (that is, the cluster formation efficiency is 100\%), then in the absence of a metallicity gradient and for a metallicity-independent $\beta(\rg)$ we expect a lower GC specific frequency for metal-poor GCs. Here we define specific frequency as the fraction of the stellar mass that is in clusters at present. However, the observed specific frequency is a factor of $\sim5$ higher at $\feh\simeq-1.5$ compared to $\feh\simeq-0.5$ \citep[in NGC~5128,][]{2002AJ....123.3108H}. 
Because of the metallicity gradient, metal-rich clusters are closer to the Galaxy centre where the tides are stronger, resulting in more mass loss for metal-rich clusters. Depending on the metallicity gradient, this effect may overcome the lower $|\dot{M}|$ at high $\feh$. 
We quantify this with our adopted implementation of $\feh$-gradient: we assume metal-rich GCs are within  $\rg\lesssim3\,\kpc$ and metal-poor GCs at $\rg\gtrsim3\,\kpc$ (see Fig.~\ref{fig:fehrg}). From this we find that the specific frequency of metal-poor(metal-rich) GCs is $\sim4\%(2\%)$. 
So we recover the same sign as the observation, but the slope is not as steep (factor of $\sim2$ vs. $\sim5$). Note that this is computed from the ratio of surviving clusters over disrupted clusters, not considering the actual field stars in the Milky Way.

%________________
\subsection{Initial density}
\label{ssec:rhohn}

Our model assumes a fixed $\rhoration = \rhohn/\rhohf = 30$ (or $\rhn/\rjeff\simeq0.05$, see Table~\ref{tab:rho}). Here we discuss the implication for the distribution of initial densities of the GC population. Because $\rhohf$ depends on the tidal field, from the orbit distribution we can derive the implied initial distribution of $\rhohn$.
For the SIS, $\rhohf$ can be found from $\omegatid$ as $\rhohf \simeq78.3\omegatid^2/G$. 
To obtain a well sampled density distribution we draw $2\times10^7$ initial masses and orbits from $f_0(\Mps,\rgvec,\vgvec)$ (equation~\ref{eq:f0}) with parameters of Model (8), restricted to  $\Mps>10^4\,\msun$, because lower mass GCs almost all dissolve. We evolve the initial masses to present-day masses with equation~(\ref{eq:Mt}), which results in $\sim2.5\times10^5$ surviving clusters. We note that although our model only evolves initial masses after stellar evolution (that is, $\Mps$), the $\dot{M}$ parameters that we use are based on initial densities before stellar evolution so we can find $\rhohn$ for each GC from $\rhohn = 30\,\rhohf(\rg,\vgvec)$.

\begin{figure}
\includegraphics[width=\columnwidth]{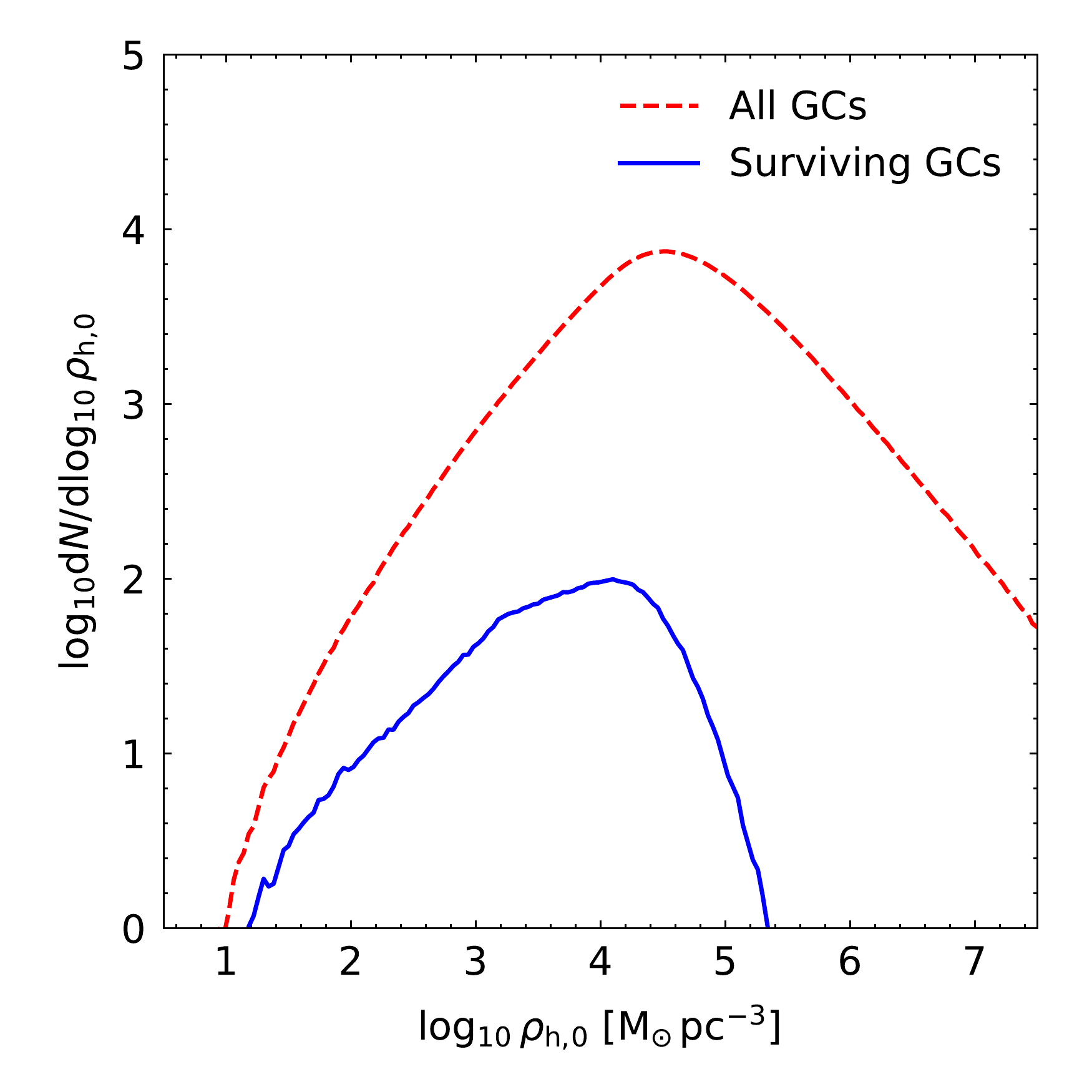}
\vspace{-8mm}
\caption{Initial density distribution for all clusters in Model (8) (red, dashed line) and for the surviving clusters (blue, solid line).}
\label{fig:rhodist}
\end{figure}

In Fig.~\ref{fig:rhodist} we plot the initial half-mass density distribution of all clusters and of the surviving clusters. Including clusters in the range $10^2-10^4\,\msun$ would increase the distribution of all GCs by a factor of $\sim110$.
This shows a peak at $\sim10^{4.5}\,\msun\,\pc^{-3}$. This is roughly an order of magnitude higher than young massive clusters in the Local Universe \citep{2010ARA&A..48..431P,2021MNRAS.508.5935B}, but it is expected that GCs at a redshift of $z\simeq4$ form denser because galaxies had higher gas fractions and velocity dispersion leading to higher pressure.
Interestingly, it was recently shown \citep[][]{2023MNRAS.522..466A} that an initial density of $\gtrsim10^4\,\msun\,\pc^{-3}$ is  what is needed to create sufficient numbers of (hierarchical) BH mergers to explain the gravitational wave sources with large primary masses ($\gtrsim20\,\msun$).

Fig.~\ref{fig:rhodist} shows that the surviving clusters have slightly lower initial densities, because the densest clusters are typically located near the Galactic centre where tidal disruption is most efficient. 

We also compare these densities to present-day (half-mass) densities of Milky Way GCs. These can be estimated by assuming that mass follows light and give the median(mean) $\rhoh \sim 300(1000)\,\msun\,\pc^{-3}$, that is, more than an order of magnitude lower than the peak density in our model after stellar evolution. This is encouraging because it is expected that the densities of clusters decrease further after stellar mass loss as the result of two-body relaxation \citep{1965AnAp...28...62H, 2011MNRAS.413.2509G}.

%__________________________________________
\subsection{Black holes in present-day GCs}
In our model, all GCs undergo a BH-dominated  phase towards the end of their evolution. 
This is the result of our assumption of a constant $\rhoration$ for all clusters. 
As already discussed in Section~\ref{ssec:rhohn}, this assumption is intended to describe the average GC. 
We can also estimate the distribution of the BH mass fraction  ($\fbh$) in our model. Towards the end of the evolution, $\Mbh$ remains approximately constant (Fig.~\ref{fig:mdot_nbody}). Assuming a constant mass-loss rate in time ($y=1$), the distribution $\dr N/\dr\fbh = (\dr N/\dr M) \, |\partial M/\partial \fbh |$  {and because $\dr N/\dr M\simeq{\rm constant}$ at low masses and $\fbh \propto 1/M$ for $\Mbh\simeq{\rm constant}$, we have $\dr N/\dr\fbh\propto \fbh^{-2}$}, so we expect the majority of clusters to have $\fbh$ close to the minimum $\fbh\simeq0.02$ for clusters with $\rhoration=30$ (see Fig.~\ref{fig:mdot_nbody}).

Various studies have pointed out that populations of stellar-mass BHs may be present in GCs,  based on their large core radii \citep{2007MNRAS.379L..40M, 2008MNRAS.386...65M}; the absence of mass segregation of stars in some GCs \citep{2016MNRAS.462.2333P, 2016ApJ...833..252A, 2020ApJ...898..162W}; the central mass-to-light ratio  \citep[for the cases of Omega Centauri and 47 Tucanae see][]{2019MNRAS.482.4713Z,2019MNRAS.488.5340B, 2019MNRAS.483.1400H}; the core over half-light radius \citep{2018MNRAS.478.1844A, 2020ApJS..247...48K} and the presence of tidal tails \citep[see][for the case of Palomar 5]{2021NatAs...5..957G}. \citet{2013MNRAS.432.2779B} suggest that all GCs apart from the ones that are classified as `core collapsed' possess BHs, which implies that 80\% of Milky Way GCs still contains BHs. 
Several studies that try to quantify $\fbh$ for larger numbers of GCs with different methods have recently become available \citep{2018MNRAS.478.1844A, 2020ApJ...898..162W, 2023arXiv230301637D}. There is generally poor agreement for individual GCs, but all studies find typical $\fbh\simeq0.01$, with exceptions like Omega Centauri \citep[$\fbh\simeq0.05$,][]{2019MNRAS.482.4713Z, 2019MNRAS.488.5340B} and Pal 5 \citep[$\fbh\simeq0.2$,][]{2021NatAs...5..957G}.

Observationally inferred $\fbh$ are very uncertain and with the available data we can only say at this moment that observations support our assumption that the majority of GCs retained some BHs until the present day. In a future modelling exercise that includes also the evolution of cluster radii it would be interesting to see how a spread in $\rhoration$ affects the final distribution of $\fbh$ to check, for example, whether the fraction of core collapsed clusters can be reproduced. 

%_______________________________________________
\subsection{Tidal perturbations and black holes} 
Several studies have invoked tidal perturbations with gas clouds in the early Universe to explain the shape of the GCMF \citep{2010ApJ...712L.184E,2015MNRAS.454.1658K,pfeffer_etal18,2018MNRAS.481.2851R}. The magnitude of this disruption mechanism is highly uncertain because it relies on poorly understood conditions in the early Universe (initial cluster densities, gas properties, etc.). The population models of \citet{pfeffer_etal18} and \citet{2018MNRAS.481.2851R} adopt initial radii of $\sim3\,\pc$ (after stellar evolution), implying densities of $\rhohn\sim10^3\,\msun\pc^{-3}$, i.e. more than an order of magnitudes lower than the peak of the initial density distribution in our models (Fig.~\ref{fig:rhodist}). Because the disruption timescale due to tidal shocks is directly proportional to the initial density, a higher initial density would decrease the disruption rate in their models. Similarly, including tidal shocks in our model would have only a small effect on our results.

The interplay between BH heating and tidal shocks is likely non-linear, but we can speculate what would happen if both effects play a role. The BHs sink to the cluster centre on a time-scale of $\sim10\,\myr$, while interactions with gas clouds can be important for up to $\sim1\,\gyr$, hence tidal shocks will predominantly remove stars from the cluster and the BH population is shielded, thereby increasing $\fbh$. So mass loss as a result of tidal shocks  amplifies the effect of BHs at later times because of an increase in $\fbh$.

\subsection{Application to other galaxies}

We applied our modelling to the Milky Way GC system, so it is interesting to consider to what extent our results apply to other galaxies. For a power-law initial GCMF with index $-2$, relaxation-driven evaporation predicts a correlation between the turnover mass, $\mto$, and the average tidal field strength experienced by the GCs, $\langle\omegatid\rangle$. One may therefore expect $\mto$ to depend on the galaxy mass/luminosity. \citet{2007ApJS..171..101J} fit `evolved Schechter functions'\footnote{These functions correspond to our $\psi(M,\omegatid,t)$, for $x=y=1$.} to luminosity functions of GC systems in early type galaxies in the Virgo Cluster. They find that $\mto$ is within a factor of $\sim2$ constant over two orders of magnitude of galaxy luminosity ($\Lgal$), with the faintest galaxies having on average a lower $\mto$.
Using galaxy scaling relations and assuming that the half-light radius of a galaxy is a proxy for the typical orbital radii of GCs, \citet{2007ApJS..171..101J} show that for galaxies with $M_B>-18$ (that is,  approximately 1 mag {fainter than} the Milky Way), the average tidal field strength depends on galaxy luminosity as $\langle\omegatid\rangle\propto L_{\rm Gal}^{0.35}$, while for brighter galaxies it goes as $\langle\omegatid\rangle\propto L_{\rm Gal}^{-0.5}$. They argue that this `peaked' relation between $\langle\omegatid\rangle$  and $L_{\rm Gal}$ is one of the explanations for the near constant $\mto$. The authors also show that in addition to $\langle\omegatid\rangle$, variations of $\Ms$ with $\Lgal$  affect the relation $\mto(\Lgal)$. For  bright galaxies they find a correlation between $\Ms$ and $\Lgal$, which offsets the anti-correlation between $\langle\omegatid\rangle$ and $\Lgal$, leading to a near constant $\mto$ and a correlation between the width of the GCMF and $\Lgal$, as is observed. These arguments apply to our suggested mass-loss recipe, with {the additional effects of $y$ and metallicity discussed in this work.}  This helps in reaching relatively high $\mto$ even in the smallest galaxies.

In the Local Group we have even fainter galaxies, with GC systems that resemble the Milky Way GCs. For example, the Fornax dSph galaxy has five old GCs with an average logarithmic mass $\mulog\simeq5.0\pm0.2$ and dispersion $\siglog\simeq0.5\pm0.2$ \citep[based on the luminosities from][and  a mass-to-light ratio of $1.8$]{2012A&A...544L..14L}.
Four of these GCs are metal-poor ($\feh\lesssim-2$), so BHs are expected to be important for their evolution and $\dot{M}$. Adopting $\Vc=20\,\kms$ for Fornax dSph, and $\rgeff=1\,\kpc$ for the GC orbits and the same model parameters as for our population model (that is, $x=0.67, y=1.33$, $\mdotref=-45\,\msun/\myr$), we find $\mulog=4.8$ and $\siglog=0.5$, in satisfactory agreement with the observed GCMF. An important constraint for GC evolution models comes from  the field stars. \citet{2012A&A...544L..14L} find that about 20\%-30\% of  all metal-poor stars ($\feh\lesssim-2$) in the galaxy resides in the four metal-poor GCs. For the simple model GCMF discussed here, and the assumption that all stars formed in GCs, we find this fraction to be $\sim20\%(40\%)$ for $\Mlo=10^2\,\msun(10^4\,\msun)$ \citep[see also][who reach a similar conclusion]{chen_gnedin23}. We therefore conclude that even in the faintest galaxies our proposed mass-loss model can reproduce the shape of the GCMF.

%%%%%%%%%%%%%%%%%%%%%%%%%%%%%%%%%%%%%%%%%%%%%%%
\section{Conclusions}
\label{sec:conclusions}

We find that two-body relaxation in a static tidal field can be the dominant disruption process in shaping the GCMF if GCs retain some of their BHs. 
Earlier studies on the effect of evaporation on the shape of the GCMF showed that models of clusters without BHs cannot reproduce the observed shape of the GCMF and its insensitivity to Galactocentric radius \citep{1998MNRAS.299.1019V, 1998A&A...330..480B}. In particular, these models find a turnover mass that is too low at large Galactocentric radii. Using $N$-body models of clusters with BHs, we show that the initial density is a critical parameter in setting the dynamical retention of BHs, and that high density clusters (relative to the tidal density) eject all their BHs and have similar $\dot{M}$ to clusters without BH. In models where dynamical BH retention is modest, the resulting $|\dot{M}|$ is still an order of magnitude higher than for models without BHs (Fig.~\ref{fig:mdot_fbh}) and the resulting mass evolution is sufficient to explain the observed turnover mass of $\sim10^5\,\msun$ at $\rg\gtrsim10\,\kpc$ (Fig.~\ref{fig:gcmfall}). 

Several additional ingredients are needed to reduce the turnover mass in the inner galaxy to a similar value. We show that the anisotropy profile of GC orbits, a Schechter-like truncation in the initial GCMF, the metallicity gradient of GCs and the effect of the past tidal evolution all reduce the decline of the turnover mass with $\rg$, with the combined effect providing a satisfactory match to the properties of Milky Way GCs (Figs.~\ref{fig:gcmf_problem1} and \ref{fig:gcmf_problem2}). {The proposed solution to the GCMF problem implies that the turnover mass gradually decreases with redshift, which is different from models that rely on early disruption mechanisms which leads to a redshift independent turnover mass. The difference may be observable with future thirty-meter class telescope and/or the James Webb Space Telescope  \citep[JWST,][]{2015MNRAS.454.1658K}. }

We present a modified analytical model for the cluster disruption rate that accounts for the effect of BHs. It is given by equation~(\ref{eq:mdot}) and for our parameters it reads
\begin{equation}
\dot{M} = -45\,\msunmyr\, \left(\frac{M}{\Mps}\right)^{-1/3} \left(\frac{\Mps}{2\times10^5\,\msun}\right)^{1/3}\frac{\omegatid}{0.32\,\myr^{-1}}.
\end{equation}
The scaling with the initial mass is the same as of clusters without BHs (equation~\ref{eq:bm03}, Fig.~\ref{fig:tdis}). The scaling with the remaining mass fraction depends on the cluster density, which sets the dynamical BH retention, and the index can be between 1/3 (high density, all BHs ejected) and $-1$ (low density, almost all BHs retained, Fig.~\ref{fig:mdot_nbody} and Table~\ref{tab:rho}). For negative indices, we obtain the `jumping' evolution of mass with time, where the mass loss rate accelerates near cluster dissolution (Fig.~\ref{fig:gcmf0}).

Although we have focused on the Milky Way system, the physical ingredients of our model are found in all galaxies and we therefore expect that the model presented here can also explain the near universality of the GCMF among different galaxies. More work is needed to confirm this.

\section*{Acknowledgements}
{We thank the referee, Douglas Heggie, for carefully reading the manuscript and providing useful feedback.}
We thank Yingtian Chen for providing results of the hierarchical model described in Appendix~\ref{sec:hierarchical} and Nate Bastian for helpful comments on the manuscript.
MG acknowledges support from the Ministry of Science and Innovation (EUR2020-112157, PID2021-125485NB-C22, CEX2019-000918-M funded by MCIN/AEI/10.13039/501100011033) and from AGAUR (SGR-2021-01069).
OG was supported in part by the U.S. National Science Foundation through grant AST-1909063 and by NASA through contract NAS5-26555 for STScI program HST-AR-16614.
Most of the processing of the results has been done using the {\sc python} programming language and the following open source modules: {\sc numpy}\footnote{http://www.numpy.org}, {\sc scipy}\footnote{http://www.scipy.org}, {\sc  matplotlib}\footnote{http://matplotlib.sourceforge.net}.

\section*{Data availability}
$N$-body data are available from MG upon reasonable request. All observational data of Milky Way GCs are from \citet{2010arXiv1012.3224H}. The {\sc Python} code {\sc EvGcmf} to evolve the GCMF is available from  \url{https://github.com/mgieles/evgcmf}.
The data used in the Appendix are available from OG  upon reasonable request.

%%%%%%%%%%%%%%%%%%%%%%%%%%%%%%%%%%%%%%%%%%%%%%
%\begin{thebibliography}{108}
%\bibliographystyle{mnras}
%\bibliography{total,gc}

\appendix

%__________________________
\section{Cluster properties from a full formation model}
\label{sec:hierarchical}

\begin{figure}
\includegraphics[width=\columnwidth]{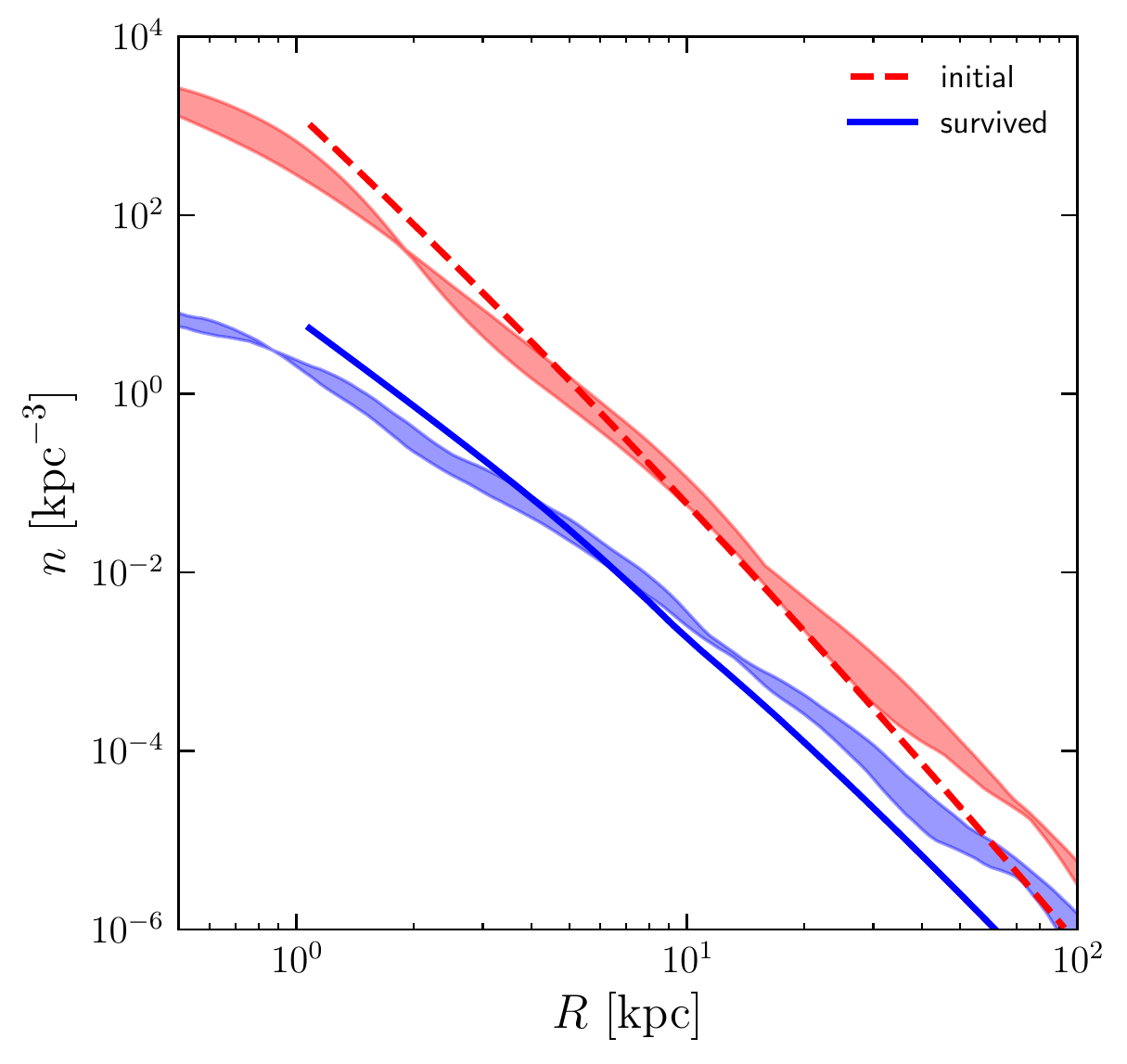}
\vspace{-5mm}
\caption{Number density profile of GCs from the mock catalog of the hierarchical formation model \citep{chen_gnedin22}. Upper red shaded region shows the range of initial densities in the three model realizations of Milky Way-like systems. Lower blue shaded region shows the range of the density of survived clusters at present. Dashed red and solid blue lines show the corresponding densities in our population Model (8), also plotted in the right column of Fig.~\ref{fig:gcmf_problem1}.}
\label{fig:rgdist2}
\end{figure}

\begin{figure}
\includegraphics[width=\columnwidth]{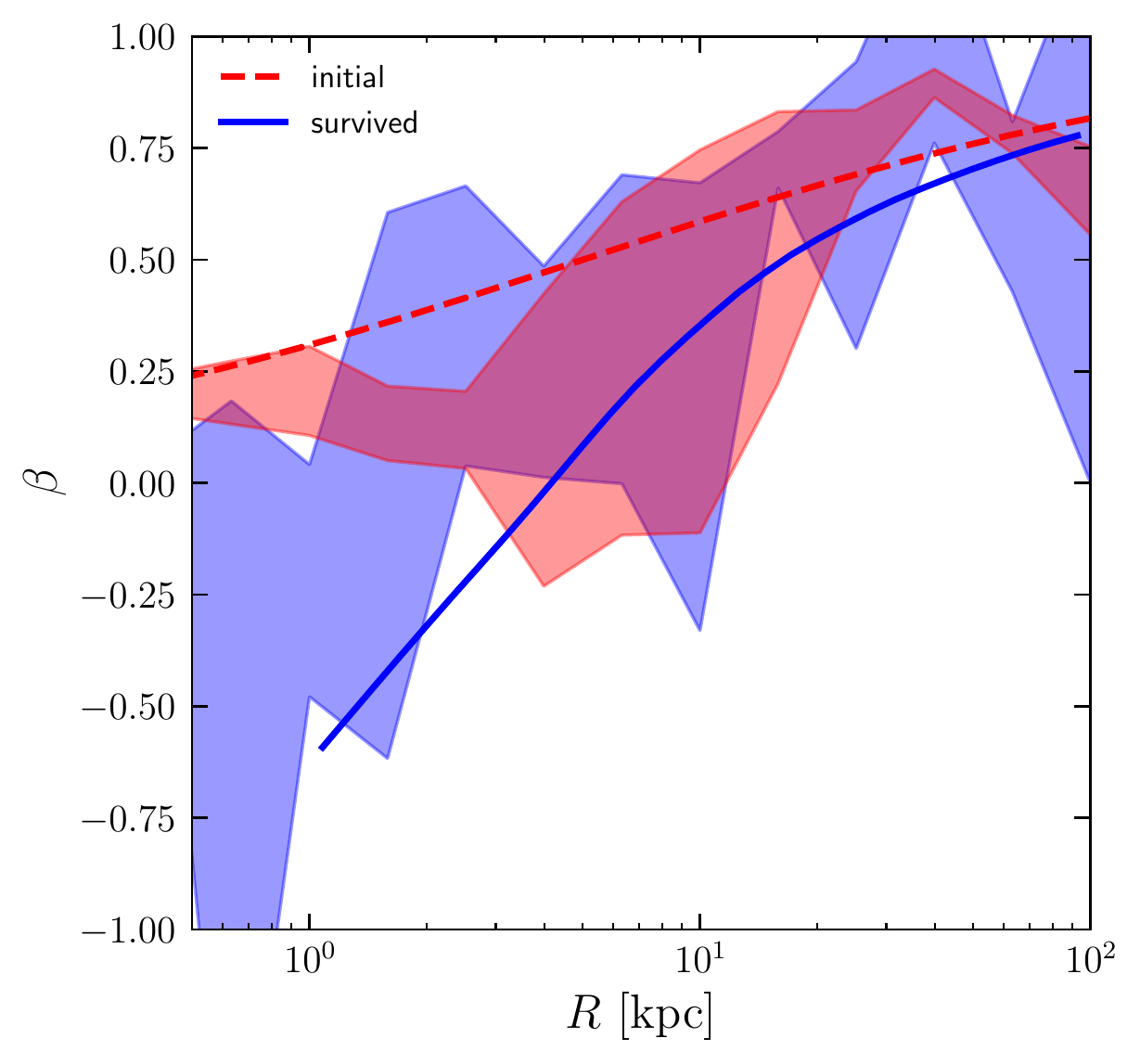}
\vspace{-5mm}
\caption{Velocity anisotropy profile of GCs from the mock catalog of the hierarchical formation model. Shaded regions show the range covered by the three realizations: red for initial, blue for final. Dashed red and solid blue lines show the corresponding densities in our population Model (8), also plotted in right column of Fig.~\ref{fig:gcmf_problem1}.}
\label{fig:beta2}
\end{figure}

{Several recent studies focused on modelling GCs from their formation in high-redshift galaxies through evolution until the present in the context of hierarchical galaxy formation \citep{choksi_etal18,pfeffer_etal18,kruijssen_etal19a,2023MNRAS.521..124R}. These studies assume that proto-GCs form in giant molecular clouds with the same initial cluster mass function as young clusters in the nearby universe. This formation process continues as long as the specific adopted criteria are satisfied and results in a range of GC formation times, typically $1-4$~Gyr after the Big Bang. Then cluster mass loss is calculated using various combinations of two-body relaxation, tidal shocks, and stellar mass loss. Despite differences in detailed implementation of all these processes, these studies reach similar conclusions that the resulting present-day GC populations can match observed properties of GC systems such as the age-metallicity distribution and the spatial and kinematic distributions. However, the resulting GCMF is usually skewed towards lower masses than observed. Given the successes in reproducing other GC properties, which support the main assumptions of the hierarchical models, the issues with the GCMF are likely to be due to inaccuracies in modelling cluster disruption. Possible inaccuracies can arise from insufficient resolution of the tidal field along cluster trajectories or using simplified mass loss prescriptions.}

{Our work can help improve the modelling of cluster evolution in hierarchical models. And in turn, hierarchical models can help test some of the assumptions made in this work.}
\cite{chen_gnedin22} presented the latest version of a GC formation and evolution model, which matches most observed properties of the Galactic GC system. Here we use the predicted properties of GCs from the model to validate our adopted initial conditions.

We use the catalog of model GC properties available online at \url{https://github.com/ognedin/gc_model_mw}. The catalog contains three systems chosen specifically to be analogous to the Milky Way in its present halo and stellar mass and in its history of the mass assembly. We use all three systems to represent a range of possible initial conditions of the Galactic GC system. Fig.~\ref{fig:rgdist2} shows the number density profile of all clusters that formed in the model as a function of the distance to the main galaxy center. This includes clusters formed throughout the cosmic time, although the middle half of them formed at the cosmic times between 11.1 to 12.5~Gyr, which is close to the assumed fixed age of 12~Gyr in our population model.

The assumed initial number density profile in our population model (equation~\ref{eq:n0}) is a good match to the range of profiles shown in red. The hierarchical model predicts a slightly shallower slope at large radii ($R > 30$~kpc) but those radii correspond to the locations of satellite galaxies in which outer GCs formed. The satellites may bring their GC systems closer to the main galaxy by dynamical friction and the eventual radii of these halo GCs would be smaller. In the population model we do not include changes of orbits due to dynamical friction, and therefore survived clusters would retain their initial radii. Thus we conclude that the hierarchical model provides support for our assumed $n_0(R)$.

To check consistency with observations, in the lower blue shaded region we show the range of number density profiles of survived clusters. 
Here the slope is more noticeably shallower than in our population model, but the difference is expected because the \cite{chen_gnedin22} model used a different GC disruption prescription with $x=y=2/3$ and $\mdotref=-45\,\msun\,\myr^{-1}$ (in our notation). The smaller $y$ (compared to $y=4/3$ in our Model 8) leads to slower disruption of low-mass clusters and allows them to survive longer at large radii where the disruption time is longer than the age. In our population model such clusters are more easily disrupted and the outer density profile steepens.

Fig.~\ref{fig:beta2} compares the velocity anisotropy profiles of the two models. Variations among the three hierarchical model realizations are large but the overall trend of initial $\beta$ increasing with radius is in reasonable agreement with the assumed form in equation~(\ref{eq:beta}). For the survived clusters, the $\beta$-profiles are even closer and both clearly predict a mildly tangential anisotropy in the inner few kpc. Thus we can conclude that the results of the full hierarchical formation model support our assumed initial conditions.

\begin{figure}
\includegraphics[width=\columnwidth]{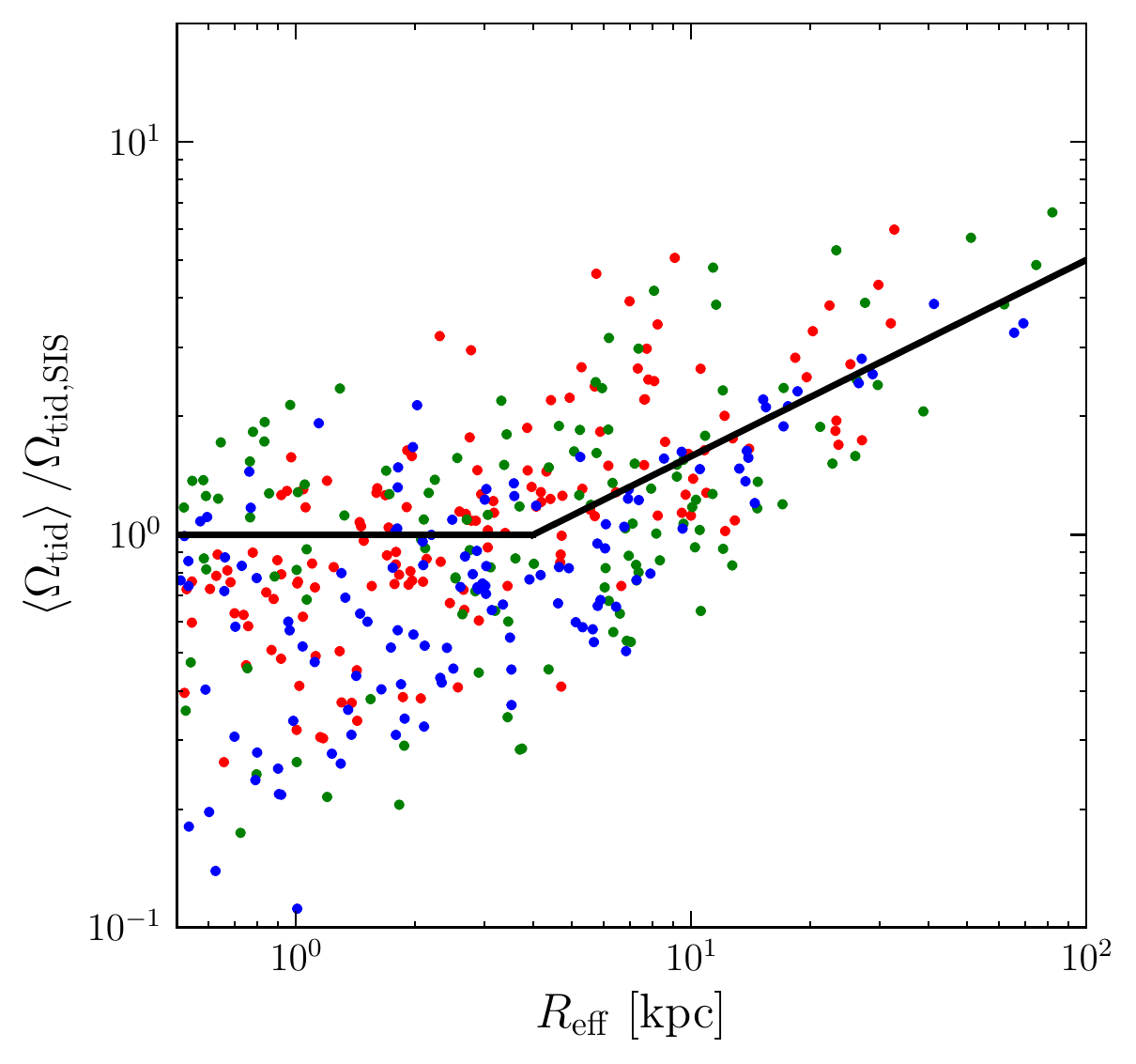}
\vspace{-5mm}
\caption{The effective strength of tidal field averaged over cluster history in the hierarchical formation model, relative to that of the best-fitting fixed SIS potential. Points show $\omegatid$ averaged over all simulation outputs for three Milky Way analog systems indicated by different colour. The solid line shows our modified expression $\Omega_{\rm tid,1}$ at $\rgeff > 4$~kpc.}
\label{fig:omegatid}
\end{figure}

The hierarchical model also allows us to investigate the evolution of the tidal field along the trajectories of model GCs. For example, \cite{meng_gnedin22} showed that a typical effective strength of tidal field $\omegatid$ was a factor of 10 higher in the first $\sim300$~Myr after cluster formation compared to the values inferred from the present-day potential. Young clusters experience stronger tides because they are still surrounded by dense gaseous and stellar structure. The tidal field can also vary rapidly in time depending on the GC trajectories. A higher fraction of GCs migrate outward from the galaxy center than inward, also leading to the weaker tidal field at present. To account for this "past evolution" of the tidal field, we calculate the time averaged $\omegatid$ experienced by survived clusters in the \cite{chen_gnedin22} model and compare it with the value in our assumed potential. Fig.~\ref{fig:omegatid} shows the {ratio of the two} for model clusters as a function of their effective radius $\rgeff = \Rp(1+\epsilon)$, where $\Rp$ is the pericentre distance of the orbit near the present. {The ratio is based on the following calculation}.

\cite{chen_gnedin22} calculated the tidal strength via a combination of the highest and lowest eigenvalues of the tidal tensor that accounts for the {tidal and} centrifugal forces: $\omegatid^2 \simeq \lambda_1-\lambda_3$. Typically, $\lambda_1>0$ and $\lambda_3<0$. For a SIS, $\lambda_1 = -\lambda_3 = \Vc^2/R^2$, and therefore
\begin{equation}
  \Omega_{\rm tid,SIS}^2(R) = \frac{2 \Vc^2}{R^2}.
\end{equation}
For a general power-law density distribution $\rho\propto R^{-n}$ with $0<n<3$
\begin{equation}
  \Omega_{{\rm tid},n}^2(R) = 4\pi G\rho(R)\, \frac{n}{3-n}
  = \frac{n\, \Vc(R)^2}{R^2},
\end{equation}
where $\Vc(R)^2 \equiv GM(R)/R$ {and $n=2$ for the SIS}.

The SIS model is a good approximation to the total mass density {in the three model realizations} in the range of radii from 1 to 100 kpc; in the inner 1 kpc the density profile approaches a core. In the middle part of the galaxy, at $\rgeff \la 4$~kpc, the SIS potential gives overall correct scaling of $\left<\omegatid\right>$ with radius, however the scatter of individual points is significant. At larger radii, the present-day $\Omega_{\rm tid,SIS}$ visibly underestimates the past tidal strength. We can approximately correct this underestimate by switching to $\Omega_{{\rm tid},n}$ corresponding to shallower distribution with $n\approx 1$. This results in a stronger tidal field at $\rgeff > 4$~kpc relative to our SIS model by a factor
\begin{equation}
  \frac{\Omega_{{\rm tid},1}^2(\rgeff)}{\Omega_{{\rm tid},2}^2(\rgeff)} = \frac{\rgeff}{4\,\mathrm{kpc}}.
  \label{eq:past}
\end{equation}
{This adopted modified expression for $\left<\omegatid\right>$ is shown by the broken line in Fig.~\ref{fig:omegatid} and is used in Models (6) and (8) to include the effect of the past evolution of GCs.}

% Don't change these lines
\bsp	% typesetting comment
\label{lastpage}
\end{document}